\documentclass[runningheads]{llncs}
\PassOptionsToPackage{obeyspaces}{url}
\usepackage{graphicx}

\usepackage{float, enumerate, xcolor, algorithm, tabto, balance, amsmath, graphicx, epstopdf, hyperref, multirow, dblfloatfix, textcomp, listings, eurosym, lstautogobble, multirow, url, tikz} 
\usepackage{amsfonts}

\usepackage[ruled,linesnumbered,algo2e]{algorithm2e}
\usepackage[table]{colortbl}
\definecolor{cellFillingColor}{rgb}{0.8, 0.8, 0.8}
\usepackage{amssymb}
\usepackage{booktabs}
\usepackage{url}
\usepackage{amssymb}
\usepackage[numbers,sort&compress]{natbib}
\usepackage[flushleft]{threeparttable}
\usetikzlibrary{arrows,backgrounds,positioning}
\usepackage{chngcntr}
\usepackage[overload]{empheq}
\usepackage{soul}
\usepackage[title]{appendix}
\usepackage{subfigure}

\usepackage{soul, listings, color}

\definecolor{dkgreen}{rgb}{0,0.6,0}
\definecolor{gray}{rgb}{0.5,0.5,0.5}
\definecolor{mauve}{rgb}{0.58,0,0.82}

\usepackage{chngcntr}
\AtBeginDocument{\counterwithin{lstlisting}{section}}

\usepackage{array}
\newcolumntype{L}{>{\arraybackslash}m{.45\columnwidth}|}

\usepackage{url}

\usepackage{pifont}
\usetikzlibrary{arrows,automata}
\usepackage[misc]{ifsym}
\graphicspath{ {./images/} }

\begin{document}
	\title{KeTS: Kernel-based Trust Segmentation against Model Poisoning Attacks}
	\titlerunning{KeTS}
	\author{
		Ankit Gangwal$^1$
		\and
		Mauro Conti$^2$
		\and
		Tommaso Pauselli$^3$\textsuperscript{(\Letter)}
	}
	\authorrunning{A.~Gangwal et al.}
	\institute{$^1$IIIT Hyderabad, India, $^2$University of Padua, Italy, $^3$Politecnico di Milano, Italy\\
		\email{gangwal@iiit.ac.in},
		\email{mauro.conti@unipd.it},
		\email{tommaso.pauselli@mail.polimi.it}
	}
	\maketitle
	
	\begin{abstract}
		Federated Learning~(FL) enables multiple users to collaboratively train a global model in a distributed manner without revealing their personal data. However, FL remains vulnerable to model poisoning attacks, where malicious actors inject crafted updates to compromise the global model's accuracy. We propose a novel defense mechanism, Kernel-based Trust Segmentation~(\textit{KeTS}), to counter model poisoning attacks. Unlike existing approaches, \textit{KeTS} analyzes the evolution of each client's updates and effectively segments malicious clients using Kernel Density Estimation~(KDE), even in the presence of benign outliers. We thoroughly evaluate \textit{KeTS}'s performance against the six most effective model poisoning attacks~(i.e., \textit{Trim-Attack}, \textit{Krum-Attack}, \textit{Min-Max} attack, \textit{Min-Sum} attack, and their variants) on four different datasets~(i.e., MNIST,  Fashion-MNIST, CIFAR-10, and KDD-CUP-1999) and compare its performance with three classical robust schemes~(i.e., \textit{Krum}, \textit{Trim-Mean}, and \textit{Median}) and a state-of-the-art defense~(i.e., \textit{FLTrust}). Our results show that \textit{KeTS} outperforms the existing defenses in every attack setting; beating the best-performing defense by an overall average of $>24$\%~(on MNIST), $>14$\%~(on Fashion-MNIST), $>9$\%~(on CIFAR-10), $>11$\%~(on KDD-CUP-1999). A series of further experiments~(varying poisoning approaches, attacker population, etc.) reveal the consistent and superior performance of \textit{KeTS} under diverse conditions. \textit{KeTS} is a practical solution as it satisfies all three defense objectives~(i.e., fidelity, robustness, and efficiency) without imposing additional overhead on the clients. Finally, we also discuss a simple, yet effective extension to KeTS to handle consistent-untargeted~(e.g., \textit{sign-flipping}) attacks as well as targeted attacks~(e.g., \textit{label-flipping}).
	\end{abstract}

	\section{Introduction}
	\label{Introduction}
	Federated Learning~(FL) is a crucial paradigm for machine learning on resource-constrained devices~(such as mobile, edge, and IoT devices)~\cite{mcmahan2017communication, bagdasaryan2020backdoor, mcmahan2016federated}. Big tech companies such as Google have already adopted it~(e.g., for the next-word prediction in Android's Gboard~\cite{mcmahan2017federated}). FL also has profound applications in healthcare for handling sensitive data~\cite{huang2019patient}. Given its potential to safeguard private and proprietary client data, FL holds significant promise, especially in light of emerging privacy regulations, e.g., GDPR. In FL, multiple devices with privacy-sensitive data collaborate to optimize a machine learning model under the guidance of a central server, while keeping the data decentralized and private. 
	\par
	The process of FL typically consists of three main steps that are repeated iteratively. First, the service provider's server distributes the current global model to the clients or a selected subset of them. Second, each client then adapts the global model using its own local training data to create a local model and sends the updates back to the server. Third, the server aggregates the local model updates according to a specified rule to generate a global model update, which is used to update the global model. A widely used FL method for aggregation in non-adversarial settings is \textit{FedAvg}~\cite{mcmahan2017communication}, which was developed by Google. This method computes the average of local model updates, weighted by the sizes of the local training datasets, to produce the global model update.
	\par
	Given the decentralized nature of FL, it is vulnerable to adversarial attacks from malicious clients. These clients may either be artificially introduced by attackers or are legitimate clients that have been compromised. As a result, FL faces the risk of various model poisoning attacks~\cite{bhagoji2019analyzing, guerraoui2018hidden}, where attackers submit malicious updates that can undermine the global model. An attacker can compromise the global model either by poisoning the local dataset~(\textit{data poisoning attacks}~\cite{biggio2012poisoning, nelson2008exploiting}) or by directly manipulating the model updates~(\textit{local model poisoning attacks}~\cite{fang2020local, bhagoji2019analyzing, bagdasaryan2020backdoor, nelson2008exploiting}). When the attacker's goal is to indiscriminately degrade the global model's predictions across a wide range of test samples, it is called an untargeted attack~\cite{fang2020local}. Conversely, when the objective is to cause misclassification for a specific class label, it is called a targeted attack~\cite{bagdasaryan2020backdoor, bhagoji2019analyzing}. Several robust aggregation rules have been proposed to address model poisoning attacks, e.g., \textit{Krum}~\cite{blanchard2017machine}, \textit{Trim-Mean}~\cite{yin2018byzantine}, and \textit{Median}~\cite{yin2018byzantine}, which rely on statistical analyses. To the best of our knowledge, \textit{FLTrust}~\cite{cao2020fltrust}, \textit{RECESS}~\cite{yan2024recess},  \textit{DnC}~\cite{shejwalkar2021manipulating} are the state-of-the-art solutions against model poisoning attacks in FL. Due to the unavailability of code, we do not consider \textit{DnC} and \textit{RECESS} in our evaluation.
	\par
	\textit{Motivation:} 
	Existing classical defense schemes~\cite{blanchard2017machine, yin2018byzantine} that rely on statistical analyses are vulnerable to optimization-based model poisoning attacks~\cite{shejwalkar2021manipulating, fang2020local}, resulting in a decline in global model accuracy. The primary issue with these techniques is the non-Independent and non-Identically Distributed~(non-IID) nature of datasets across clients. The state-of-the-art defense, \textit{FLTrust}, is also vulnerable in highly non-IID environments because its root dataset diverges significantly from clients' local data distributions. As a result, benign client updates are frequently misclassified as statistical outliers. Therefore, a solution is missing in the literature that can distinguish malicious clients from benign clients in highly non-IID environments and in the presence of benign outliers.
	\par
	\textit{Contributions:} The major contributions of our work are as follows:
		\par 
		1. We propose Kernel-based Trust Segmentation~(\textit{KeTS}), a novel defense mechanism against model poisoning attacks. \textit{KeTS} analyzes each client's updates to compute an individual trust score that takes into account their historical contributions. \textit{KeTS} segments the trust scores via Kernel Density Estimation~(KDE)~\cite{parzen1962estimation} to distinguish between benign and malicious clients effectively, even in the presence of benign outliers.
		\par 
		2. We empirically evaluate \textit{KeTS} against six untargeted model poisoning attacks in a white-box scenario, which is the most difficult combination to defend against. We perform our evaluations using both image and tabular datasets (i.e., MNIST~\cite{lecun1998mnist}, Fashion-MNIST~\cite{xiao2017fashion}, CIFAR-10~\cite{krizhevsky2009learning}, and KDD-CUP-1999~\cite{kdd_cup_1999_data_130}). Our results show that \textit{KeTS} not only outperforms the existing defenses, but it also achieves all three defense objectives~(i.e., fidelity, robustness, and efficiency).
		\par 
		3. In a series of further experiments, we (i)~analyze the behavior of different clients with KeTS; assess the impact of (ii)~changing the degree of non-IID partitions, (iii)~attacker population, (iv)~poisoning approaches, (v)~number of local epochs; and finally, (vi)~we propose a simple extension to KeTS to handle consistent-untargeted as well as targeted attacks.
	\par
	\textit{Organization:} This paper is organized as follows. \S~\ref{related works} reports the background and key concepts in FL. \S~\ref{Threat model} presents our threat model. \S~\ref{section:KeTS} elucidates our proposed approach. \S~\ref{section:eval} presents our evaluation results. Finally, \S~\ref{conclusion} concludes the paper.
	
	\section{Related Works}
	\label{related works}
	\subsection{Background}
	\label{sec:RLbackground}
	\textbf{Common FL framework:} \textit{FedAvg}~\cite{mcmahan2017communication} has established itself as the de facto standard for FL, which works as follows. At the beginning of each round, the server sends the global model to a randomly chosen subset of clients. These clients train the model on their local datasets over multiple local epochs. Next, the clients upload their trained models back to the server. The server then aggregates these updates by calculating a weighted average of each client's gradient. Here, the weight is proportional to the size of the client's training dataset. In contrast to traditional distributed stochastic gradient descent~(which typically uses a single epoch), \textit{FedAvg}'s use of multiple epochs significantly reduces the number of communication rounds; which makes it far more communication-efficient.
	\\
	\textbf{Non-IIDness in FL:} Non-IID data in FL is characterized by substantial differences in distribution and features across the data contributed by different client participants. In real-world scenarios, such variability stems from various interconnected factors~(including user behaviors, and data collection methods)~\cite{lu2024federated}. Non-IID scenarios are typically emulated using: (1)~\textit{label distribution skew}, where the label distributions \( P(y_i) \) vary across parties; or (2)~\textit{feature distribution skew}, where the feature distributions \( P(x_i) \) vary across parties. Nonetheless, the conditional probability \( P(y_i \mid x_i) \) remains the same~\cite{li2022federated}. We adopt the former strategy using a distribution-based label imbalance to simulate our non-IID settings. Specifically, each party is allocated a proportion of samples for each label according to a \textit{Dirichlet} distribution.
	\subsection{Poisoning attacks in FL}
	\label{sec:RLPoisoning}
	An attacker can compromise the global FL model by either modifying the local dataset~(\textit{data poisoning attacks}~\cite{biggio2012poisoning, nelson2008exploiting}) or manipulating the model updates themselves~(\textit{local model poisoning attacks}~\cite{fang2020local, bhagoji2019analyzing, bagdasaryan2020backdoor, nelson2008exploiting}). An untargeted attack~\cite{fang2020local} aims to degrade the global model's performance across a range of test samples, while a targeted attack~\cite{bagdasaryan2020backdoor, bhagoji2019analyzing} seeks to cause misclassification for a specific class.\\ 
	\textbf{Krum-attack and Trim-attack:} Fang et al.~\cite{fang2020local} introduced a general framework for local model poisoning attacks that can be tailored to any aggregation rule. Their approach formulates the attack as an optimization problem. Here, the goal is to maximize the deviation of the global model update from the original update direction, achieved by crafting poisoned gradients from malicious clients. Depending on the aggregation rule, the optimization problem takes on different forms. Specifically, they designed attacks for \textit{Krum} (termed as \textit{Krum-Attack}), as well as for \textit{Trim-Mean} and \textit{Median} (termed as \textit{Trim-Attack}).\\
	\textbf{Aggregation attacks:} The optimization problem can be refined by incorporating perturbation vectors and scaling factors~\cite{shejwalkar2021manipulating}. By embedding the product of the perturbation vector and the scaling factor into the optimization objective, the scaling factor can be iteratively adjusted to maximize the attack's effectiveness. The authors~\cite{shejwalkar2021manipulating} propose three specific attacks:  AGR-tailored, AGR-agnostic \textit{Min-Max}, and AGR-agnostic \textit{Min-Sum}. In the AGR-tailored attack, the attacker utilizes knowledge of the server's aggregation rules to design an objective function that maximizes the scaling factor while adhering to those rules. Conversely, when the aggregation rules are unknown, the AGR-agnostic attacks focus on crafting poisoned gradients that deviate significantly from the majority of benign updates while evading detection by defense mechanisms. In the \textit{Min-Max} attack, the malicious gradient is crafted so that its maximum distance from any other gradient is bounded by the maximum distance between any two benign gradients. Conversely, in the \textit{Min-Sum} attack, the malicious gradient is designed to ensure that the sum of its squared distances from all benign gradients is bounded by the sum of squared distances between any benign gradient and the rest of the benign gradients. The perturbation vector represents any direction within the upload space that, when multiplied by the scaling factor, is utilized to perturb the mean of the benign gradients. We considered two different types of perturbation gradients: (1)~inverse unit vector~(denoted as Unit-Vector), which is a unit vector pointing in the opposite direction of the mean of the benign gradients; and (2)~inverse standard deviation~(denoted Std.-Vector), which is the negative component-wise standard deviation of the benign uploads.
	\subsection{Existing Byzantine robust aggregation rules}
	\label{sec:RLDefenses}
	A summary of the existing key defenses against poisoning attacks is as follows:\\
	\textbf{Krum}~\cite{blanchard2017machine}: It is a majority-based method. Given $c$ malicious and $n$ total clients, \textit{Krum} identifies the gradient that minimizes the sum of squared distances to the $n - c - 2$ closest neighbors and selects it as the final aggregation result.\\
	\textbf{Trimmed-Mean (Trim-Mean)}~\cite{yin2018byzantine}: It is a robust aggregation method that considers each coordinate of the upload separately, making it a coordinate-wise aggregation technique. For each model parameter, the server collects and sorts its value from all clients' updates. Using a trim parameter $k$ (where $k < \frac{n}{2}$ over $n$ total clients), the server discards the $k$ smallest and $k$ largest values, then calculates the mean of the remaining $n - 2k$ values. To ensure robustness, $k$ should be at least as large as the number of attackers; i.e., the \textit{Trim-Mean} approach is effective only when the proportion of malicious clients is below~50\%.\\
	\textbf{Median}~\cite{yin2018byzantine}: It is another coordinate-wise aggregation method. For each model parameter, it sorts the values across all local model updates. Rather than using the mean value after trimming, the \textit{Median} method selects the median value of each parameter as its corresponding value in the global model update.\\
	\textbf{FLTrust}~\cite{cao2020fltrust}: The server in \textit{FLTrust} utilizes a small dataset to participate in each iteration and generate a gradient benchmark. As outlined by the authors~\cite{cao2020fltrust}, we sample the root dataset uniformly at random from the union of the clients' clean training datasets while maintaining a homogeneous distribution of samples across classes, ensuring each class contains an equal number of samples. Each local model update is normalized to have the same magnitude as the server's model update. Clients whose cosine similarity between the updates and the server's update is negative are not considered for aggregation. The remaining normalized updates are aggregated via a weighted average based on their cosine similarity. \\
	\textbf{DnC}~\cite{shejwalkar2021manipulating}: It leverages singular value decomposition based spectral methods for outliers detection and removal. DnC provides strong theoretical robustness guarantees for removing malicious updates when benign updates are \textit{iid} but, it struggles in \textit{non-iid}, white-box scenarios. \\
	\textbf{RECESS}~\cite{yan2024recess}: This proactive defense constructs test gradients to analyze client responses. Benign clients optimize toward their local data distribution, while malicious ones introduce erratic changes to maximize poisoning. It detects gradient abnormalities via directional and magnitude shifts. A key limitation of RECESS is its reliance on querying clients with constructed gradients, which can introduce significant overheads and hinder scalability in large FL systems.

	\section{Threat Model}
	\label{Threat model}
	\textbf{Attacker:} Our attacker aims to degrade the global model's accuracy without being detected. We adopt the attack settings from previous works~\cite{cao2020fltrust, yan2024recess}. Specifically, the server remains uncompromised while an attacker controls malicious clients. These malicious clients may either be fake clients injected by the attacker or genuine clients compromised by the attacker. The malicious clients can send crafted local model updates to the server in each iteration of the FL training process. We primarily focus on untargeted attacks against established aggregation schemes, as they result in greater losses in the final accuracy of the global model. Typically, the attacker has partial knowledge of the FL system, including the local training data, local model updates from the malicious clients, the loss function, and the learning rate. We consider the white-box scenario, in which the attacker can access benign clients' updates and compute an attack based on this knowledge. Such a strong adversary represents the most impactful threat~model.\\
	\textbf{Defender}: The defense is assumed to be implemented on the server side, which is unaware of the number of attackers and does not have access to the client datasets. However, the server has complete access to the global model and the local model updates from all clients during each iteration. Additionally, it gathers updates from all the clients across previous iterations.\\
	\textbf{Defense goal}: The goal is to develop an FL technique that accomplishes Byzantine resilience against harmful clients without compromising accuracy. An effective defense mechanism must achieve all the following three defense objectives~\cite{cao2020fltrust}:
		\par 
		1. \textbf{Fidelity:} When there are no adversarial attacks, the defense technique should guarantee that the classification performance of the global model is unaffected. In particular, the technique should be able to generate a global model with a classification accuracy similar to \textit{FedAvg}, assuming no attacks.
		\par
		2. \textbf{Robustness:} In the presence of malevolent clients carrying out severe poisoning attacks, the method should maintain the classification accuracy of the global model. The aim is to design a method that can train a global model amid an attack while maintaining performance comparable to a global model trained by \textit{FedAvg} in a non-adversarial situation.
		\par
		3. \textbf{Efficiency:} The approach shouldn't result in additional communication and computation costs, particularly for the clients. The clients in FL are often devices with limited resources. Thus, the goal is to create a technique that does not add to the client's workload.

	\section{Kernel-based Trust Segmentation~(\textit{KeTS})}
	\label{section:KeTS}
\begin{figure}[!htbp]
		\centering
		\vspace{-1.25em}
		\includegraphics[trim= 10mm 205mm 35mm 10mm, clip, width=0.7\textwidth]{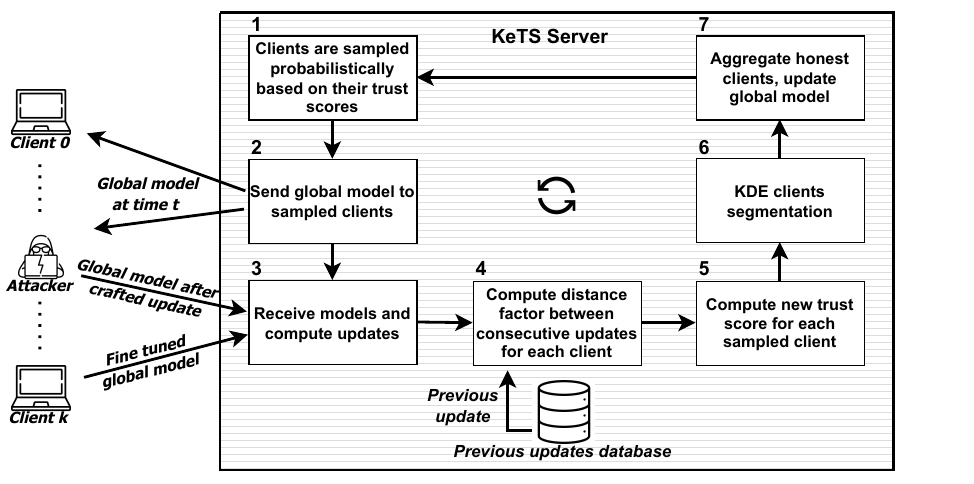}
		\vspace{-1em}
		\caption{An overview of \textit{KeTS}.}
		\vspace{-1.5em}
		\label{fig:ketsOverview}
	\end{figure}

	We propose \textit{KeTS}, a novel defense against model poisoning attacks in FL. \textit{KeTS} derives its name from its reliance on Kernel-Based methods to estimate the probability density function of clients' trust scores. \textit{KeTS}  aims to detect malicious gradient without labeling honest clients as false positives even in largely non-IID environments. The intuition behind \textit{KeTS} is that benign and malicious clients behave differently, i.e., benign clients aim to minimize a loss function related to their data while malicious clients aim to maximize the poisoning of the global model without being detected solving an optimization problem. In essence, benign gradients tend to point to their local distribution and remain consistent during the epochs. On the other hand, malicious gradients - solved on the basis of the optimization problems - change inconsistently. The same is demonstrated in \S~\ref{behavioral}, where the cosine similarity and Euclidean distance exhibit an inconsistent pattern for malicious clients. The existing defense schemes~(\textit{FLTrust}~\cite{cao2020fltrust}, \textit{Krum}~\cite{blanchard2017machine}, \textit{Trim-Mean}~\cite{yin2018byzantine}, \textit{Median}~\cite{yin2018byzantine}) rely on statistical analyses of the uploaded values or on utilizing similarities/differences among various uploads/previous global update. In contrast, we focus on analyzing each client individually and their history by calculating trust scores. Each client's trust score is based solely on its submitted uploads, which enables \textit{KeTS} to function effectively in heterogeneous scenarios; where updates may vary significantly due to a non-IID distribution of class labels across clients. \figurename~\ref{fig:ketsOverview} presents an overview of \textit{KeTS}.
	\par
	We now present the following related to our scheme: general setting in \S~\ref{generalsetting}, reputation score and penalty in \S~\ref{reputationScore}, trust score in \S~\ref{tscores}, segmentation in \S~\ref{segmentation}, and then aggregation in in \S~\ref{aggregation}.	
	
	\subsection{General setting}
	\label{generalsetting}
	The server sends a global model to the sampled clients at each global epoch. The sampling method is based on trust scores, in fact the higher the trust score, the higher the probability of being sampled. At the outset, all trust scores are initialized to 1. Therefore, in the first global round, we sample all the clients to quickly observe variations in the trust scores, allowing us to start detecting potential attackers and avoid aggregating malicious updates. Each client will fine-tune its local model on its local dataset for a certain number of local epochs and send the resulting model updated to the server. To prevent any additional calculations on the client side, the server computes the resulting update \(u\) for each client \(i\) at a given epoch \(t\), as defined by:
	\begin{equation}
		u_{i}^{t} = {Local\_model}_{i}^{t} - {Global\_model}_{i}^{t-1},
	\end{equation}
	where each client \(i\) fine-tunes its local model at each local training iteration \(k\) according to:
	\begin{equation}
		{LocalModel}_{i}^{k+1} = {LocalModel}_{i}^{k} - \eta \cdot \nabla \mathcal{L}_{i},
	\end{equation}
	where $\eta$ is the fixed learning rate for all the clients and $\nabla \mathcal{L}$ is local Loss function.
	Once the sampled clients send their updates, the server collects and stores them.
	\subsection{Reputation score and penalty}
	\label{reputationScore}
	We define the updates as \( \boldsymbol{u} = \{ u_0, u_1, u_2, \dots, u_n \} \), where each update corresponds to a client's local update.
	After computing the updates for each client, the server calculates a reputation score by analyzing two measures of abnormality: \textit{direction} and \textit{magnitude}.
	\par
	The attacker can control the direction of malicious updates in order to deviate from the original direction of the global gradient. As noted in \S~\ref{behavioral}, the upload direction of benign clients points to its distribution of the local dataset since it minimizes its local loss function and is consistent with the upload of the previous global epoch.
	We use \textit{cosine similarity}~($S_i$) as a metric to detect the angular change in direction between the upload of client \( i \) at time \( t \)~(denoted as \( \mathbf{u}^{t}_{i} \).) and the update at time \( t-1 \)~(denoted as \( \mathbf{u}^{t-1}_{i} \)).
	\begin{equation}
		\label{Eq:CS}
		S_i \left( u^{t}_{i}, u^{t-1}_{i} \right) = \frac{u^{t}_{i} \cdot u^{t-1}_{i}}{\| u^{t}_{i} \|_2 \cdot \| u^{t-1}_{i} \|_2}.
	\end{equation}
	\par
	An attacker can also manipulate the magnitude of the poisoned gradients, particularly when larger than the benign gradients.
	Hence, we decide to leverage \({l}_{2}\) distance to measure the magnitude difference between two consecutive updates. Formally:
	\begin{equation}
		\label{Eq:ED}
		\|\mathbf{u}_{i}^{t} - \mathbf{u}_{i}^{t-1}\|_2 = \sqrt{\sum_{j} \left( u_{i}^{t,(j)} - u_{i}^{t-1,(j)} \right)^2}.
	\end{equation}
	\par
	Now, we introduce penalty. The penalty serves as a measure of dissimilarity between two consecutive updates, focusing exclusively on the updates of an individual client, without considering their distance from other clients' updates~(as in \textit{Krum}~\cite{blanchard2017machine}, \textit{Trim-Mean}~\cite{yin2018byzantine}, \textit{Median}~\cite{yin2018byzantine}) or the server's update~(as in \textit{FLTrust}~\cite{cao2020fltrust}).
	
	\begin{theorem}[Penalty] In our FL setting, the penalty \(\textit{Py}\) for the update uploaded by the \(i\)-th client can be defined as:
		\begin{equation}
			\label{Eq:DF}
			Py = 
			\begin{cases} 
				(1 - S_i \left( u^{t}_{i}, u^{t-1}_{i} \right)) + \|{u}_{i}^{t} - {u}_{i}^{t-1}\|_2 , & \text{if } S_i \left( u^{t}_{i}, u^{t-1}_{i} \right) \geq 0, \\
				\frac{{TScore}_{i}^{t-1}}{\beta} & \text{otherwise}
			\end{cases}
		\end{equation}
		where $\beta$ is a fixed-parameter that influences speed of the score shift, and ${TScore}^{t-1}$ is the trust score of i-th client at the end of the previous global round.
	\end{theorem}
	\par
	Cosine similarity is initially used to discriminate malicious updates. Clients, whose update direction deviates from the previous one and have a cosine similarity lower than $0$ are promptly discarded; and their trust score is set to $0$. It is achieved by assigning a penalty, which, when multiplied by $\beta$, results in the previous trust score, leading it to be set to 0~(cf. Eq.~\ref{tscoreupdate}). The trust score update process is described in \S~\ref{tscores}. Penalty is negatively correlated to the Cosine Similarity \({S}_{i} \) and positively correlated to the Euclidean Distance \({l}_{2}\). Using the sum between the variables, rather than the product, ensures that the penalty remains responsive to changes in both \({S}_{i} \) and \({l}_{2}\) independently.
	
	\subsection{Trust score}
	\label{tscores}
	The clients' trust scores are based on the history of each client's updates, enabling the tracking of long-term performance rather than focusing solely on the current iteration. By using trust scores, we can analyze the long-term behavior of clients; as their value is not determined by the current epoch alone, but is the cumulative result of the entire training process. Initially at $t=0$, we assign the same trust score $=$1 to each client. Then at each iteration, we update it in the following way:
	\begin{equation}
		\label{tscoreupdate}
		\text{Trust}^{t} = \max\left(0, \text{Trust}^{t-1} - {\beta} \times {Py} \right).
	\end{equation}
	\par
	The rule above prevents the trust scores to be lower than 0. In fact, when a client reaches 0, it means that it will no longer be sampled and is eliminated from the clients pool.
	
	\subsection{Segmentation}
	\label{segmentation}
	Once we have our trust scores for the sampled clients, we want to segment them in order to detect the malicious clients.
	They can be easily divided into two distinct groups. To this end, we use Kernel Density Estimation~(KDE)~\cite{parzen1962estimation}. We chose KDE for its statistical robustness in one-dimensional clustering (in our case, clustering of trust scores). One-dimensional data is inherently more structured and sortable, making KDE a statistically sound and well-suited; because it efficiently detects local minima in the density distribution - unlike multidimensional techniques that are unnecessary and less effective in such settings.
	\begin{theorem}[Kernel Density Estimation]
		Let \( X_1, X_2,\dots , X_n \in \mathbb{R}^d \) be an independent and identically distributed (IID) random sample drawn from an unknown distribution \( P \) with density function \( p \). The KDE of the probability density function \( p \) at any point \( x \) can be expressed as:
		\begin{equation}
			\hat{p}_n(x) = \frac{1}{n h^d} \sum_{i=1}^{n} K\left( \frac{x - X_i}{h} \right),
		\end{equation}
		where \( K: \mathbb{R}^d \to \mathbb{R} \) is a smooth function called the kernel function, and \( h > 0 \) is the bandwidth parameter that controls the degree of smoothing. A common kernel function is the \textbf{Gaussian Kernel}, defined as:
		\begin{equation}
			K(x) = \frac{\exp\left( -\frac{x^2}{2} \right)}{v_{1,d}},
		\end{equation}
		where \( v_{1,d} \) is given by the integral:
		\begin{equation}
			v_{1,d} = \int_{\mathbb{R}^d} \exp\left( -\frac{x^2}{2} \right) \, dx.
		\end{equation}
	\end{theorem}
	\par
	KDE smooths each data point into a continuous bump, with the shape of the bump determined by the particular kernel function K(x). Individual bumps are summed to obtain a density estimate. In regions with a high concentration of observations, the density will be higher because many bumps contribute to the estimate. Inversely, in regions with fewer observations, the density will be lower as only a small number of bumps influence the density estimate. Then we consider the local minima of the resulting density function as the boundaries for our clusters. The bandwidth \(h\) at every iteration is estimated using \(estimate\_bandwidth\) function from \textit{scikit-learn}; this function chooses the bandwidth by calculating the maximum distance between a point and its nearest neighbors for each sample in the dataset. It then averages these distances to determine the bandwidth. We note that KDE is used solely to segment a 1D vector of \textit{n} trust scores. Given \textit{m} evaluation points (1000 in our setting), KDE has a complexity of $O(nm)$ (without optimization), while \textit{bandwidth} selection requires $O(n^2)$. Since \textit{n} is bounded by the number of clients, the complexity overhead is negligible. Algorithm~A.1 in Appendix~\ref{appendix:algo} presents \textit{KeTS}' segmentation process.

	\subsection{Aggregation}
	\label{aggregation}
	Let the Kernel Density Estimator be \(\hat{p}_n(x)\), where trust scores \(x \in [0, 1]\). We denote the local minima of \(\hat{p}_n(x)\) as \(m_1, m_2, \dots, m_k\), such that \(m_1 < m_2 < \dots < m_k\), and \(m_k\) is the last local minimum before \(x = 1\). The set of trust scores higher than \(m_k\) can be written as:
	\begin{equation}
		S = \{ x \mid x > m_k, \; x \in [0, 1] \},
	\end{equation}
	where \(m_k\) is given by:
	\begin{equation}
		m_k = \max\{ x_i \mid \hat{p}_n'(x_i) = 0, \; \hat{p}_n''(x_i) > 0, \; x_i \in [0, 1] \}.
	\end{equation}
	\par
	Here, $\hat{p}_n'(x_i)$ and $\hat{p}_n''(x_i)$ denote the first and second order derivative of $\hat{p}_n(x_i)$, respectively. We consider the uploads of the selected clients as honest and aggregate them using an algorithmic equivalent of \textit{FedAvg}~\cite{mcmahan2017communication}, i.e., first computing a weighted average of the selected uploads and then adding the resulting update to the previous global model~\cite{mcmahan2017communication}. Prior works~(such as RECESS~\cite{yan2024recess} and \textit{FLTrust}~\cite{cao2020fltrust}) use a weighted aggregation mechanism based on trust scores. Their approach sacrifices the weight each client has based on its dataset cardinality, at the risk of more false positives. Moreover, selecting the clients to be aggregated allows us more flexibility in the final aggregation scheme.
	\par
	\textit{KeTS} focuses on detecting untargeted model poisoning attacks by identifying inconsistencies in client update histories. Although data poisoning attacks~(e.g., label flipping~\cite{biggio2012poisoning, fung2020limitations}) lead to consistent updates from corrupted datasets, they shall appear benign to our method. By combining \textit{KeTS} with traditional Byzantine-robust aggregation schemes~(such as \textit{Median} and \textit{Trim-Mean}), we can first detect model poisoning attacks and then address data poisoning through traditional defenses. Algorithm~A.2 in Appendix~\ref{appendix:algo} presents \textit{KeTS}' aggregation process.

	\section{Evaluation}
	\label{section:eval}
	\S~\ref{configurations} presents our evaluation setup. We thoroughly evaluate \textit{KeTS} against six different poisoning attacks and compare its performance with the existing defenses~(cf.~\S~\ref{MainResults}). We analyze the behavior of clients with \textit{KeTS}~(cf.~\S~\ref{behavioral}) and examine the effect of the degree of non-IID partitions~(cf.~\S~\ref{noniidsValues}). We further investigate the impact of attacker population~(cf.~\S~\ref{percentageOfAttackers}), poisoning approaches~(cf.~\S~\ref{posioningApproaches}), and the number of local epochs~(cf.~\S~\ref{numberOfLocalEpochs}). Finally, we extend KeTS for other types of attacks~(cf.~\S~\ref{sec:KeTSv2}). We also explore the feasibility of \textit{FLTrust},  a state-of-the-art defense, to match \textit{KeTS}'s performance~(cf.~Appendix~\ref{fltrust})
	
	\subsection{Settings}
	\label{configurations}
	We evaluate \textit{KeTS} against untargeted model poisoning attacks in FL. Our focus is primarily on non-IID settings as these scenarios pose significant challenges to traditional robust aggregation schemes due to the high number of benign client outliers. To simulate non-IID scenarios among clients, we use the \textit{Dirichlet Partitioner}~(originally from \textit{Flower}~\cite{beutel2020flower}) as in the work~\cite{yurochkin2019bayesian}. We sequentially divide the data over each label. A fraction of the data for each label is drawn for each client from \textit{Dirichlet} distribution and adjusted for balancing.
	\par
	\textit{Datasets:} All our experiments are written in Python 3.10.10 environment. We use MNIST~\cite{lecun1998mnist}, Fashion-MNIST~\cite{xiao2017fashion}, KDD-CUP-1999~\cite{kdd_cup_1999_data_130}, CIFAR-10~\cite{krizhevsky2009learning} datasets in our experiments. We set \textit{Dirichlet} concentration parameter \textbf{= 0.5} for the first three datasets and \textbf{2.0} for CIFAR-10  to simulate non-IID client partitions. For MNIST, we use a simple fully connected network (712 × 512 × 10). For Fashion-MNIST, we employ a CNN with two convolutional layers (32 and 64 filters of size 3x3), followed by ReLU activations and 2x2 max-pooling. The output is then flattened and passed through two fully connected layers (600 and 120 units) with a 25\% dropout. For KDD-CUP-1999 we used a fully connected network (41 x 128 x 23). We used a VGG~\cite{simonyan2014very} for CIFAR-10.
	\par
	\textit{Configurations:} All attacks are conducted in a white-box scenario, where the attacker has access to all clients' updates, making it the \textbf{most difficult} to defend against. We consider a fixed number of global and local epochs. We use mini-batch gradient descent; we used in CIFAR-10 a momentum of 0.9. The initial trust scores are set to~1, and the baseline $\beta$ is set to~0.1. The number of clients is fixed, with attackers accounting for 20\%. \tablename~\ref{tab:setting} shows our experiment settings.
	\begin{table}[H]
		\centering
		\vspace{-1em}
		\caption{Dataset details.}
		\label{tab:setting}
		\resizebox{\columnwidth}{!}{%
			\begin{tabular}{|c|c|c|c|c|c|c|c|c|c|c|}
				\hline
				\textbf{Dataset} & \textbf{Classes} & \textbf{Size} & \textbf{Dimension} & \textbf{\begin{tabular}[c]{@{}c@{}}\# of\\ clients\end{tabular}} & \textbf{\begin{tabular}[c]{@{}c@{}}\# of\\ selected \\ clients\end{tabular}} & \textbf{\begin{tabular}[c]{@{}c@{}}\# of\\ local \\ epochs\end{tabular}} & \textbf{\begin{tabular}[c]{@{}c@{}}\# of\\ global \\ epochs\end{tabular}} & \textbf{\begin{tabular}[c]{@{}c@{}}Batch \\ size\end{tabular}} & \textbf{\begin{tabular}[c]{@{}c@{}}Learning \\ rate\end{tabular}} & \textbf{\begin{tabular}[c]{@{}c@{}}\% of\\ attackers\end{tabular}} \\ \hline
				MNIST 0.5~\cite{lecun1998mnist} & 10 & 60,000 & 28x28x1 & 100 & 80 & 5 & 50 & 200 & 0.001 & 20 \\ \hline
				Fashion-MNIST 0.5~\cite{xiao2017fashion} & 10 & 60,000 & 28x28x1 & 100 & 80 & 5 & 50 & 128 & 0.001 & 20 \\ \hline
                CIFAR-10 2.0~\cite{krizhevsky2009learning} & 10 & 50.000 & 32x32x3 & 50 & 50 & 4 & 50 & 200 & 0.01 & 20 \\ \hline
                KDD-CUP-1999 0.5~\cite{kdd_cup_1999_data_130} & 23 & 800,000 & 41 & 100 & 80 & 1 & 30 & 128 & 0.001 & 20 \\ \hline
			\end{tabular}
			}
	\vspace{-1em}
	\end{table}
	\par
	\textit{Attacks and defenses:}
	We consider six different poisoning attacks, viz., \textit{Trim-Attack}~\cite{fang2020local}, \textit{Krum-Attack}~\cite{fang2020local}, \textit{Min-Max}~(Unit-Vector)~\cite{shejwalkar2021manipulating}, \textit{Min-Max}~(Std.-Vec-tor)~\cite{shejwalkar2021manipulating}, \textit{Min-Sum}~(Unit-Vector)~\cite{shejwalkar2021manipulating}, and \textit{Min-Sum}~(Std.-Vector)~\cite{shejwalkar2021manipulating}. As for the defenses, we consider three classical robust schemes (i.e., \textit{Krum}~\cite{blanchard2017machine}, \textit{Trim-Mean}~\cite{yin2018byzantine}, and \textit{Median}~\cite{yin2018byzantine}) and one state-of-the-art defense~(i.e., \textit{FLTrust}~\cite{cao2020fltrust}). We do not consider \textit{DnC} defense~\cite{shejwalkar2021manipulating} due to the unavailability of its code. The parameters for these defense mechanisms are configured to their default specified values. For \textit{Trim-Mean}, the parameter $k$ is configured to correspond to the number of attackers. We use accuracy as the metric to evaluate the performance of trained models~(both with and without poisoning). The decrease in accuracy indicates the severity of an attack, and a higher accuracy signifies a stronger defense. All the reported results are averaged over 10 runs.
	
	\subsection{Performance results against poisoning attacks}
	\label{MainResults}
\begin{figure}[h!]
		\vspace{-1.25em}
		\centering
		\subfigure[MNIST.]{
			\label{fig:MNIST}
			\includegraphics[trim= 0mm 0mm 55mm 0mm, clip,width=.38\textwidth]{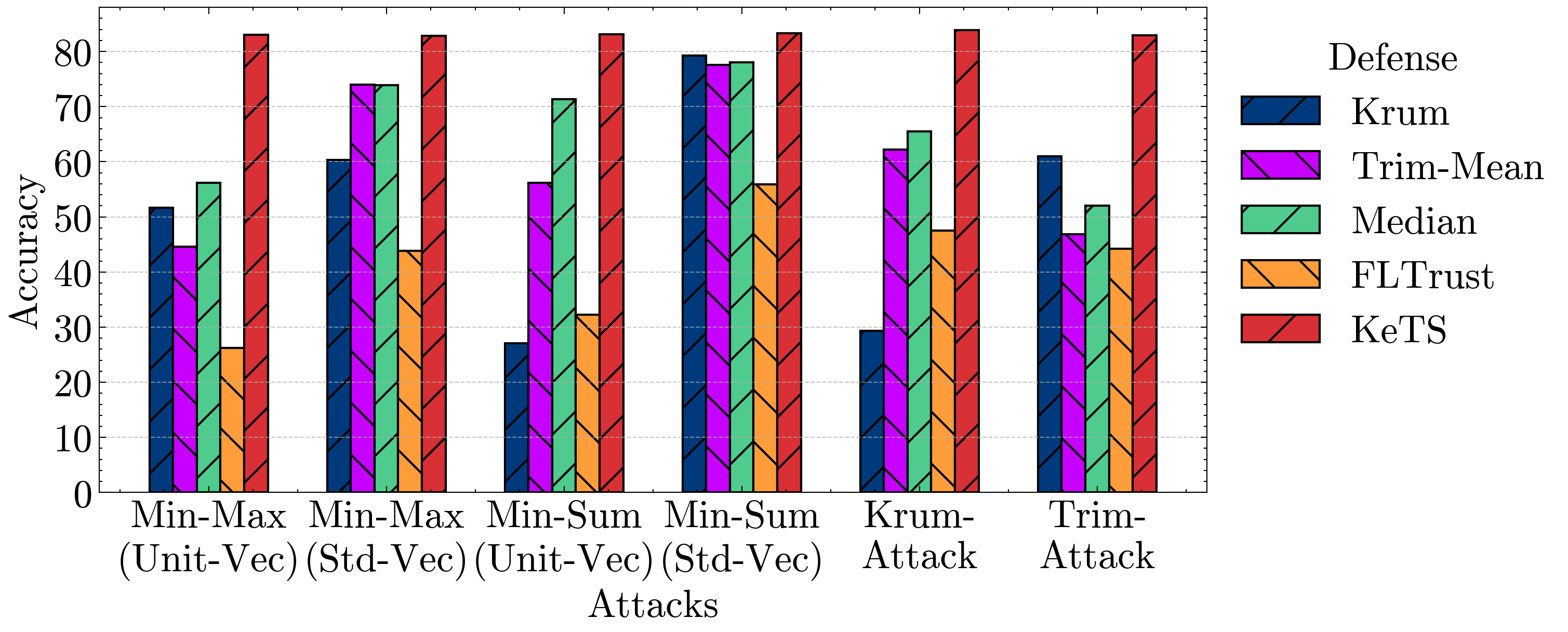}
		}
		\subfigure[Fashion-MNIST.]{
			\label{fig:fashion-MNIST}
			\includegraphics[trim= 0mm 0mm 55mm 0mm, clip, width=.38\textwidth]{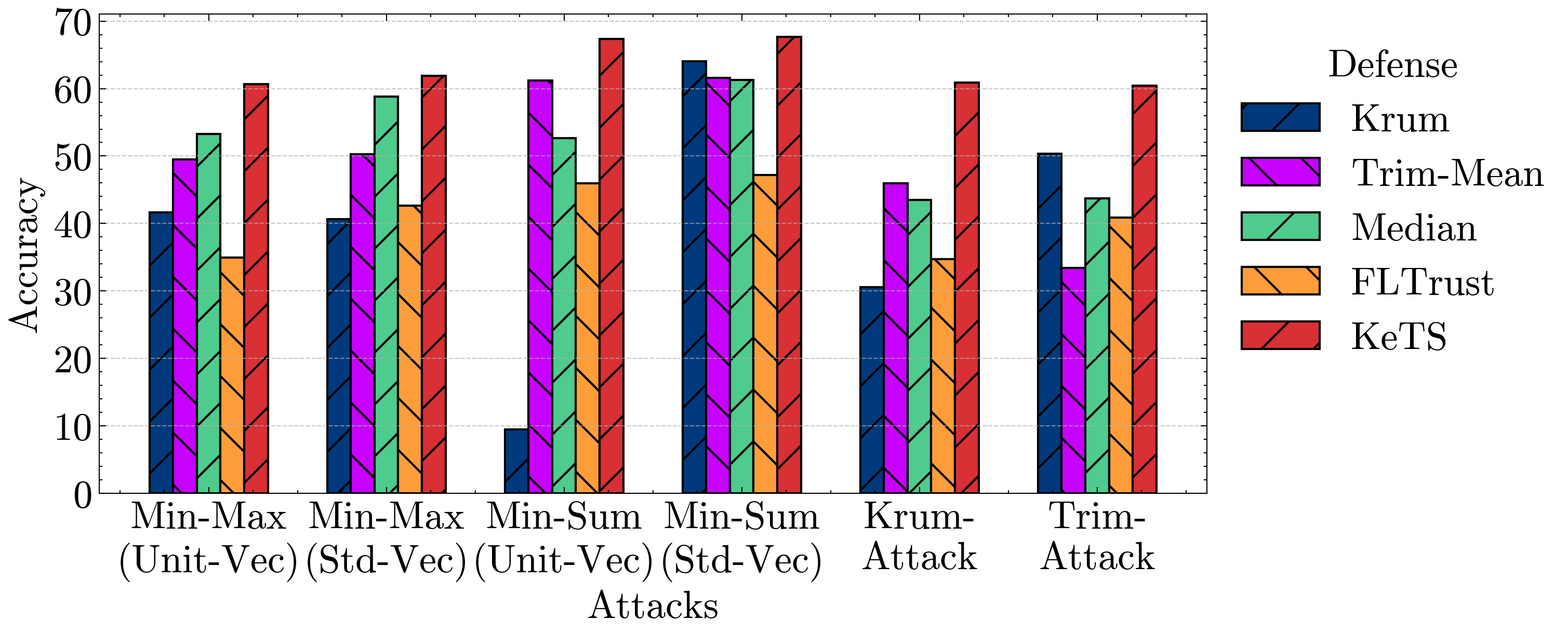}
		}
            \subfigure{
			\includegraphics[trim= 200mm 25mm 2mm 5mm, clip,width=.12\textwidth]{images/TableHisto_Fashion.png}
		}
		\addtocounter{subfigure}{-1}

		\vspace{-0.5em}
            \subfigure[KDD-cup-1999.]{
			\label{fig:KDD-cup-1999}
			\includegraphics[trim= 0mm 0mm 55mm 0mm, clip, width=.38\textwidth]{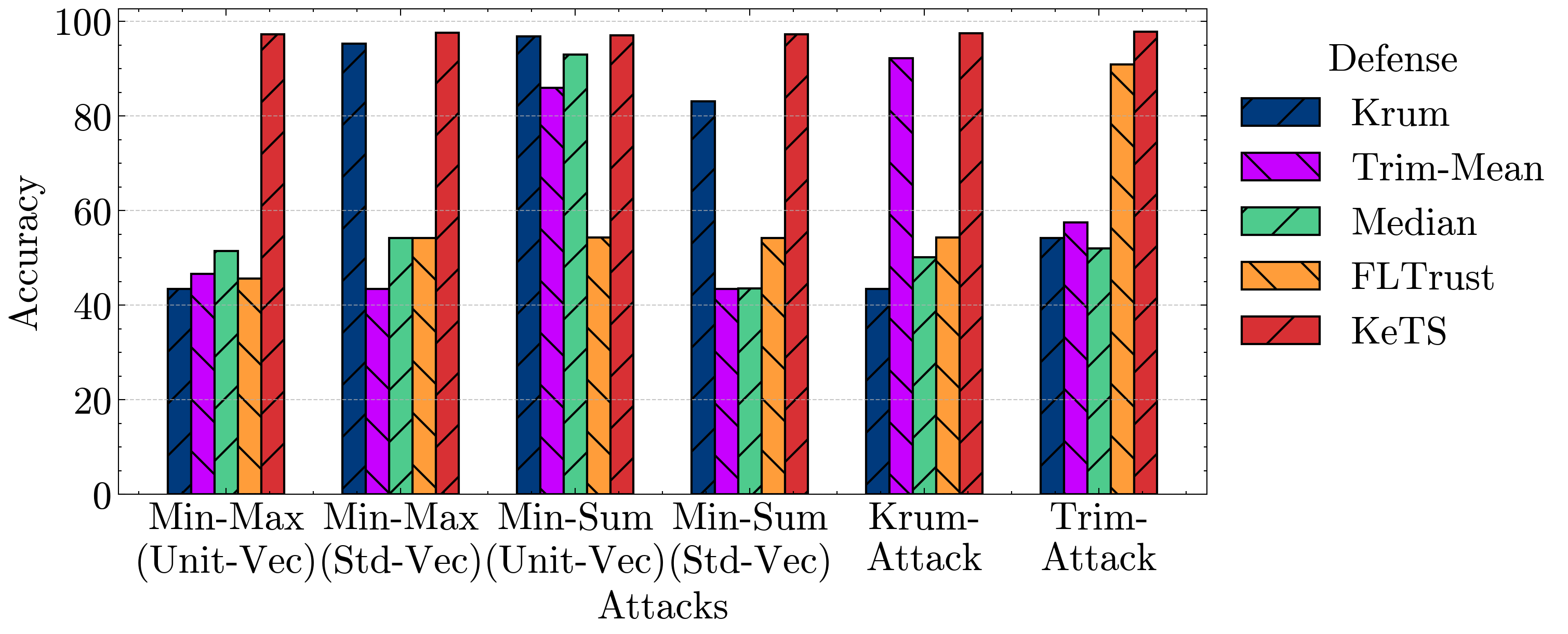}
		}
            \subfigure[CIFAR-10.]{
			\label{fig:CIFAR-10}
			\includegraphics[trim= 0mm 0mm 55mm 0mm, clip, width=.38\textwidth]{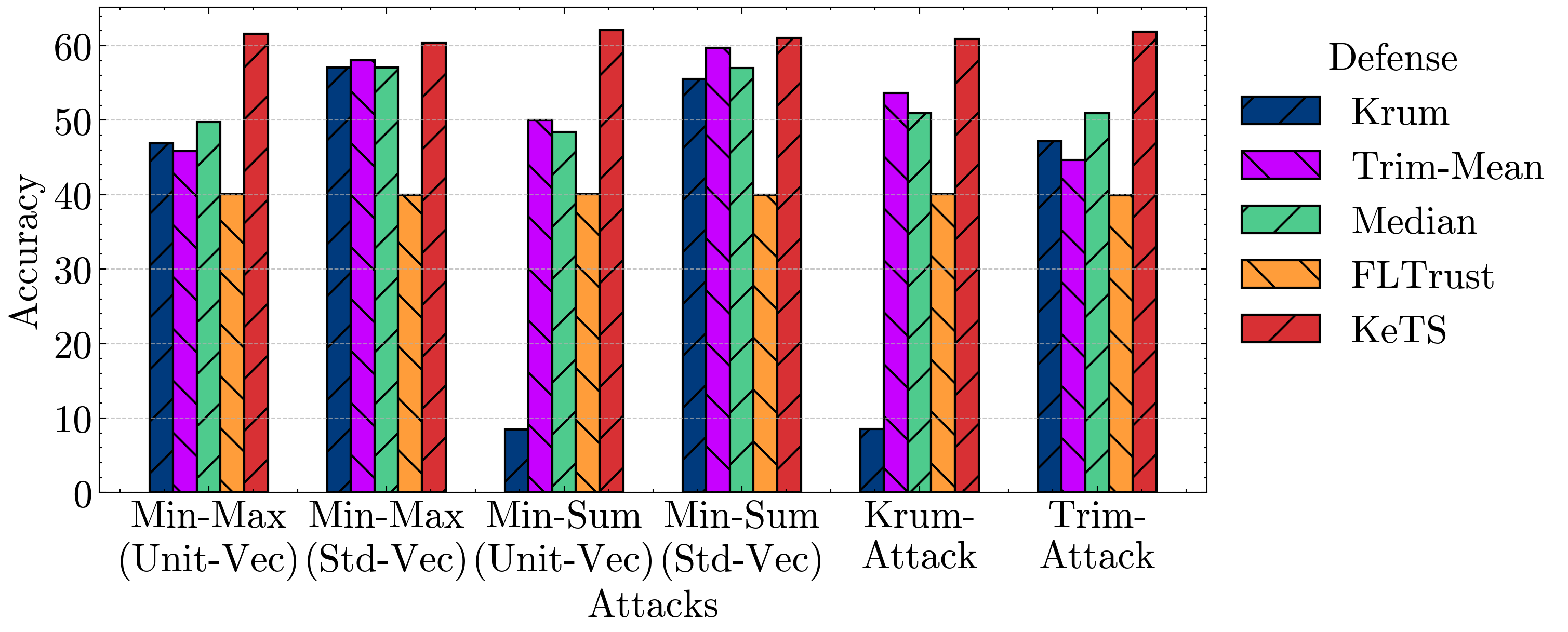}
		}~~~~~~~~~~~~~~~~
        \vspace{-1em}
		\caption{The average FL accuracy with different defense mechanisms against different poisoning attacks over all the four datasets in a white-box scenario.}
		\vspace{-1.6em}
	\label{fig:MainTableToHistogram}
\end{figure}

	Due to our \textit{Dirichlet}-based splitting method~(cf.~\S~\ref{configurations}), all the datasets exhibit a high level of non-IID in the class distribution across clients. Furthermore, once a client is considered an attacker by KeTS, its dataset is no longer available for the training process. Due to the high heterogeneity in class distribution among clients, an excluded dataset may contain the majority of samples for a particular class. In our evaluation, the excluded datasets are always the same, i.e., the attackers are consistently the same clients while evaluating different defenses. Figs.~\ref{fig:MNIST}-\ref{fig:CIFAR-10} show the average FL accuracy on the server test set for MNIST, Fashion-MNIST, KDD-cup-1999, and CIFAR-10, respectively.
\par
As shown in \figurename~\ref{fig:MainTableToHistogram}, \textit{KeTS} outperforms the existing defenses in effectively defending against all the untargeted poisoning attacks considered. Furthermore, \textit{KeTS} fulfills all the three defense objectives outlined in \S~\ref{Threat model} as follows: 
	\par 
	- \textbf{Fidelity:} We now evaluate the accuracy achieved by KeTS in the absence of attacks and compare it with the accuracy obtained by FedAvg under the same conditions. Our results demonstrate that KeTS achieves comparable performance, thereby satisfying the fidelity objective. In particular, without attacks, KeTS reaches an accuracy of 83.67\% on MNIST compared to 83.77\% with FedAvg; 71.23\% on Fashion-MNIST compared to 71.37\% with FedAvg; 97.13\% on KDD-Cup-1999 compared to 97.20\% with FedAvg; and 62.17\% on CIFAR-10 compared to 63.29\% with FedAvg. Our results confirm that KeTS does not significantly impact model performance in benign settings.
    \par 
    - \textbf{Robustness:} Unlike other defenses, \textit{KeTS} is robust because its average accuracy under any attack~(i.e., $>82$\% on MNIST, $>60$\% on Fashion-MNIST, $>97$\% on KDD-cup-1999, and $>60\%$ on CIFAR-10) clearly beating the existing defenses in every case~(cf.~\figurename~\ref{fig:MainTableToHistogram}).
	\par 
	- \textbf{Efficiency:} In \textit{KeTS}, clients incur no computational overhead while the server performs relevant basic operations~(e.g., storing each client's previous update, calculating cosine similarity and Euclidean distance, and executing segmentation using \textit{KDE}). Therefore, \textit{KeTS} is efficient and has negligible computational complexity~(cf. \S~\ref{segmentation}).\\
\textit{Discussion:} We find that the existing techniques could not defend as effectively as \textit{KeTS}, because one or both are true for the existing defenses: (1)~Benign outliers are misclassified as false positives due to the heterogeneity of client dataset partitions, which prevents their aggregation by the server. (2)~Aggregation-agnostic attacks~(i.e., \textit{Min-Max}, \textit{Min-Sum}; both variants) are designed to bypass classical aggregation rules. On the other hand, aggregation-tailored attacks~(i.e., \textit{Trim-Attack} and \textit{Krum-Attack}) are transferable to other aggregation schemes~\cite{fang2020local}, which allows malicious updates to be aggregated by the server.

\subsection{Analysis of client behavior}
\label{behavioral}
\begin{figure}[!htbp]
	\vspace{-1.4em}
	\centering
	\subfigure{
		\includegraphics[trim= 0mm 0mm 0mm 0mm, clip,width=.21\textwidth]{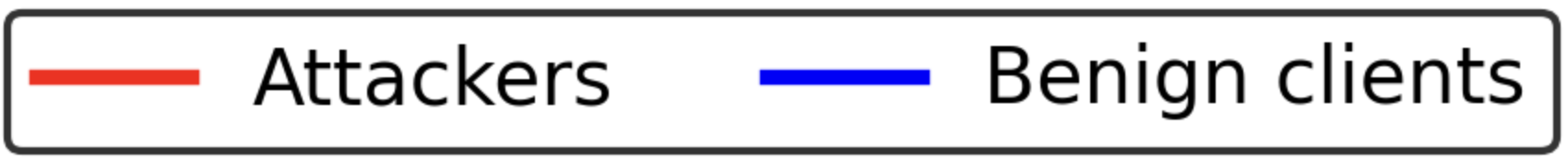}
	}
	\vspace{-1em}
	
	\subfigure{
		\label{fig:image1}
		\includegraphics[width=.21\textwidth]{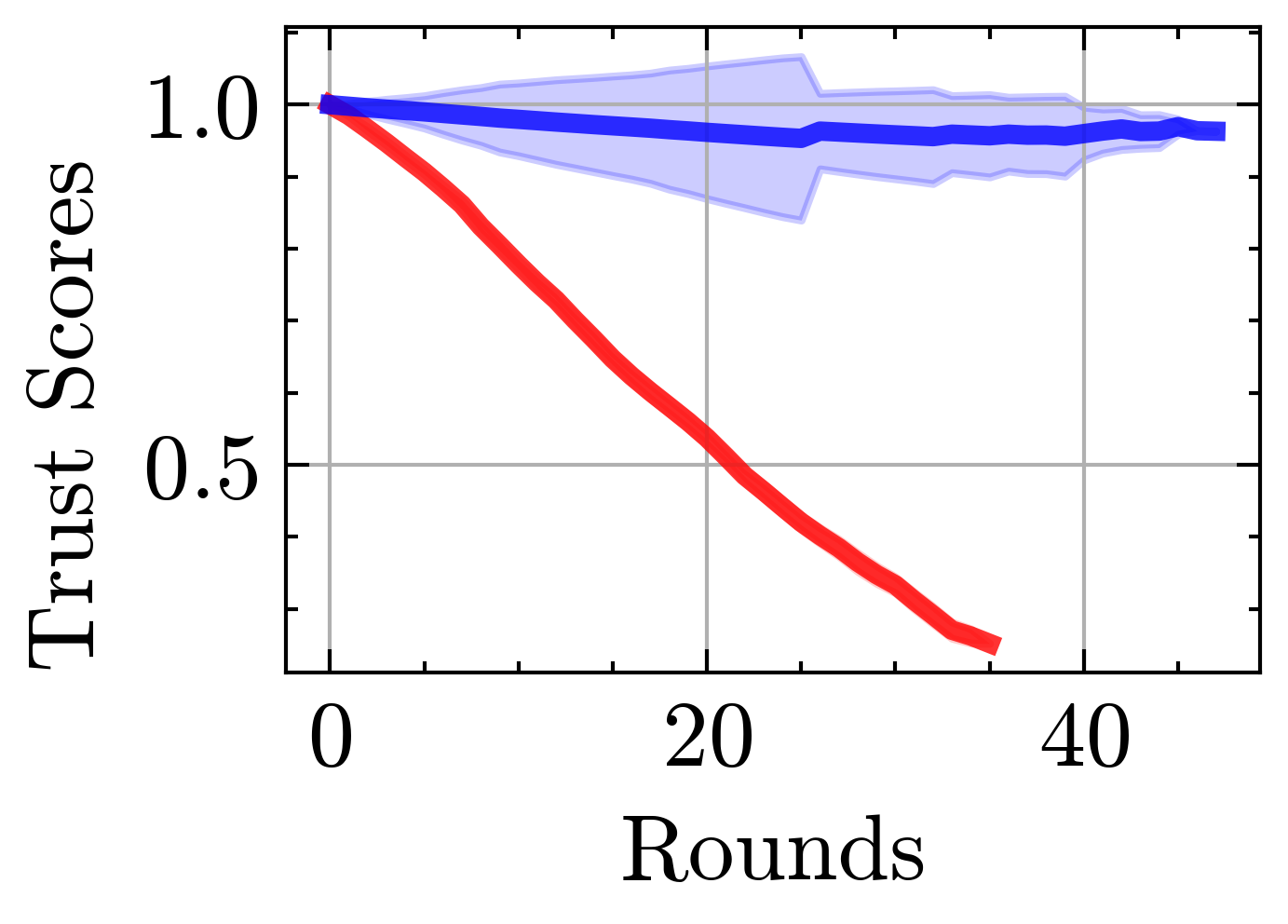}
	}
	\subfigure{
		\label{fig:image2}
		\includegraphics[width=.21\textwidth]{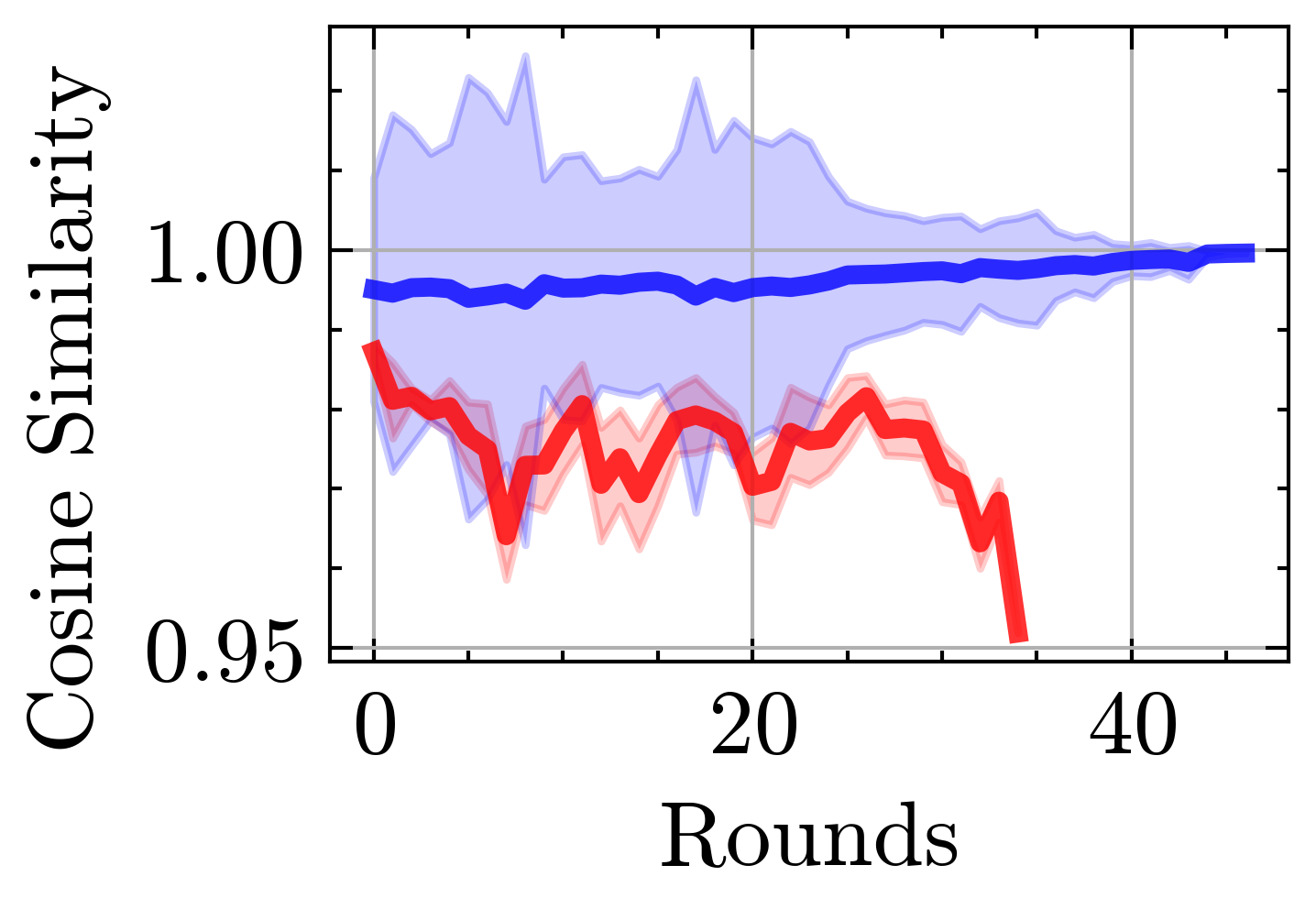}
	}
	\subfigure{
		\label{fig:image3}
		\includegraphics[width=.21\textwidth]{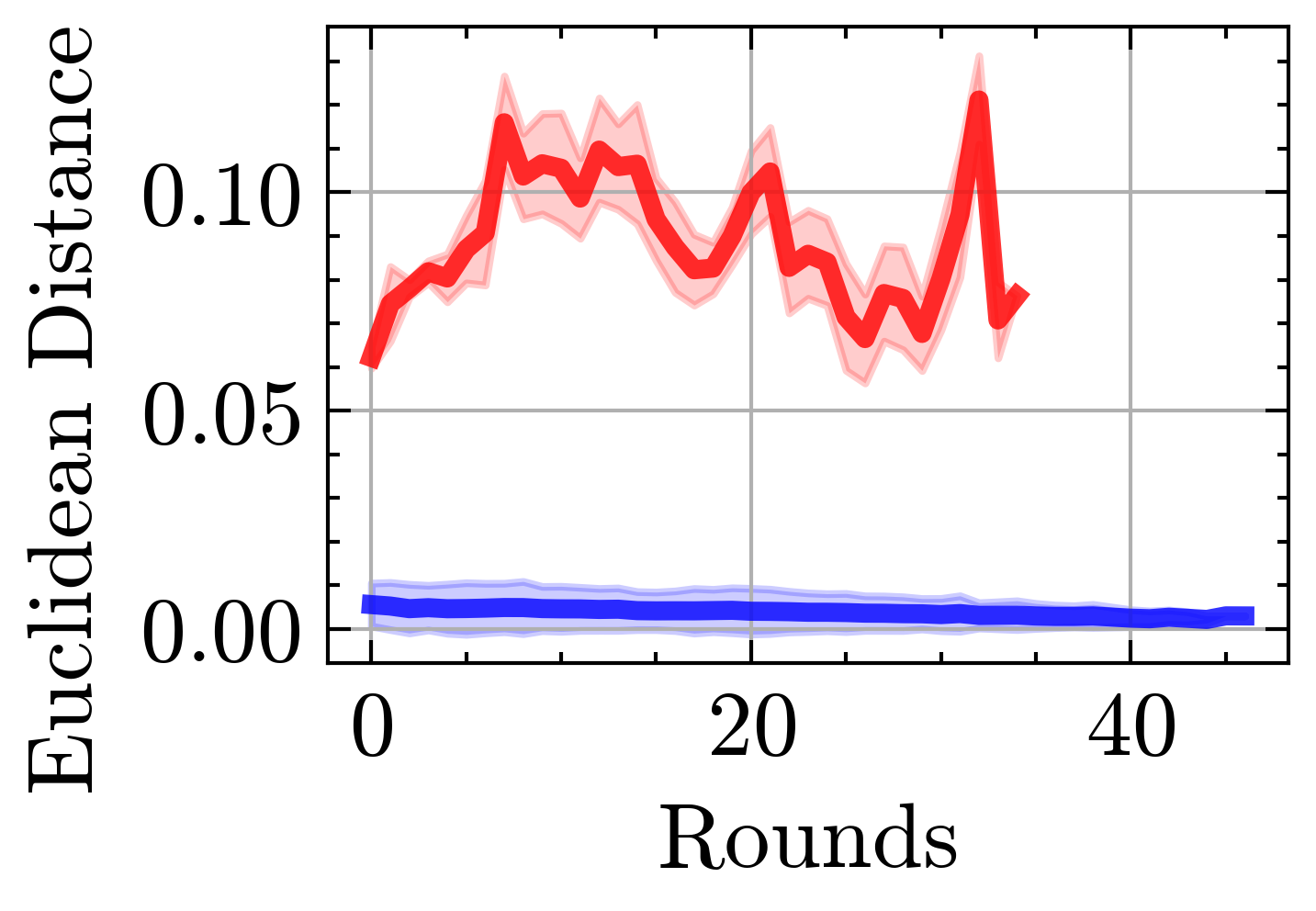}
	}
	\vspace{-1em}
	
	\addtocounter{subfigure}{-4}
	\subfigure[\textit{Min-Max}~(Unit-Vector) attack]{
		\label{fig:PerfminMaxAttack}
		\includegraphics[trim= 1500mm 0mm 0mm 0mm, clip, clip,width=.4\textwidth]{images/legendLocalBehave1.png}
	}\vspace{-0.5em}
	
	\subfigure{
		\label{fig:image4}
		\includegraphics[width=.21\textwidth]{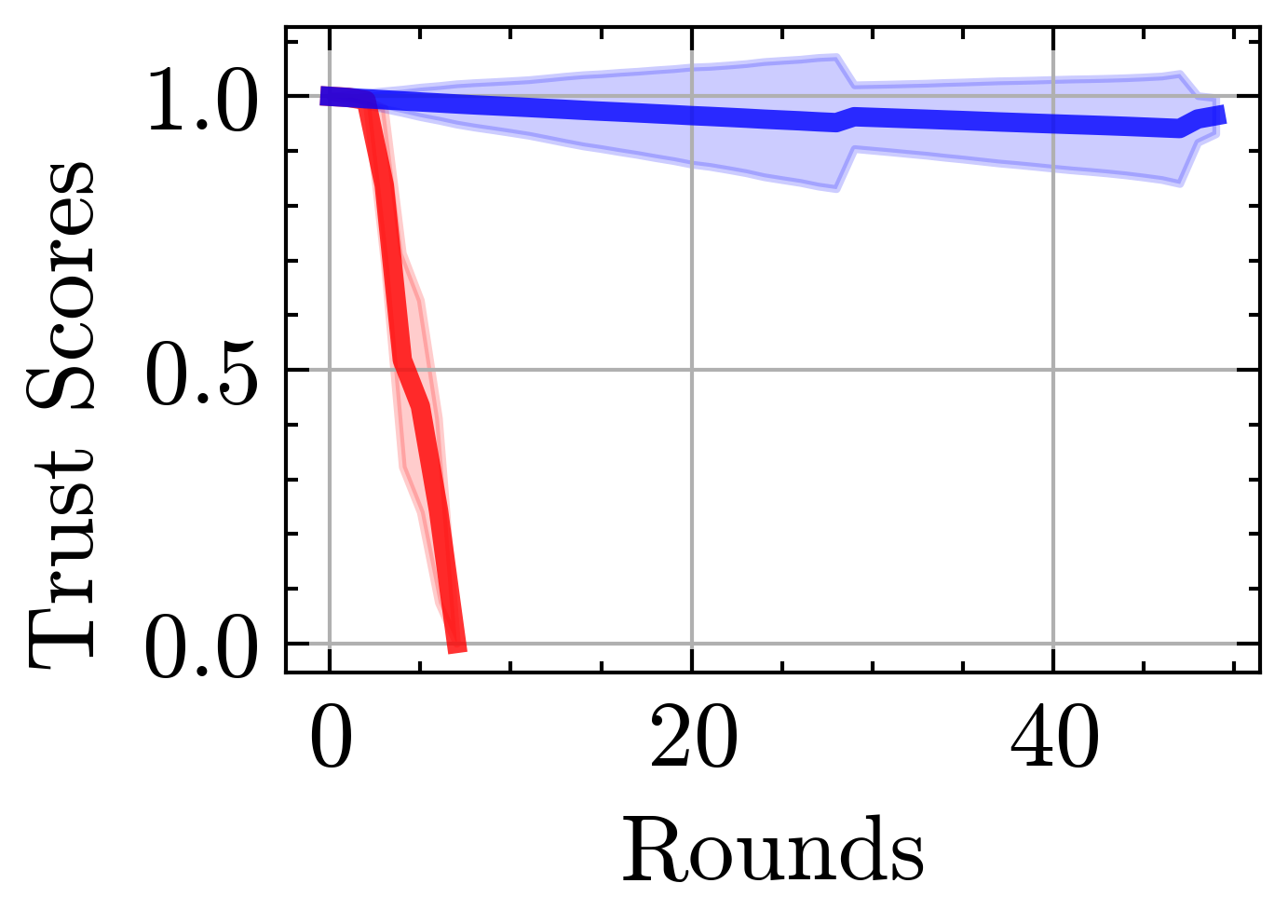}
	}
	\subfigure{
		\label{fig:image5}
		\includegraphics[width=.21\textwidth]{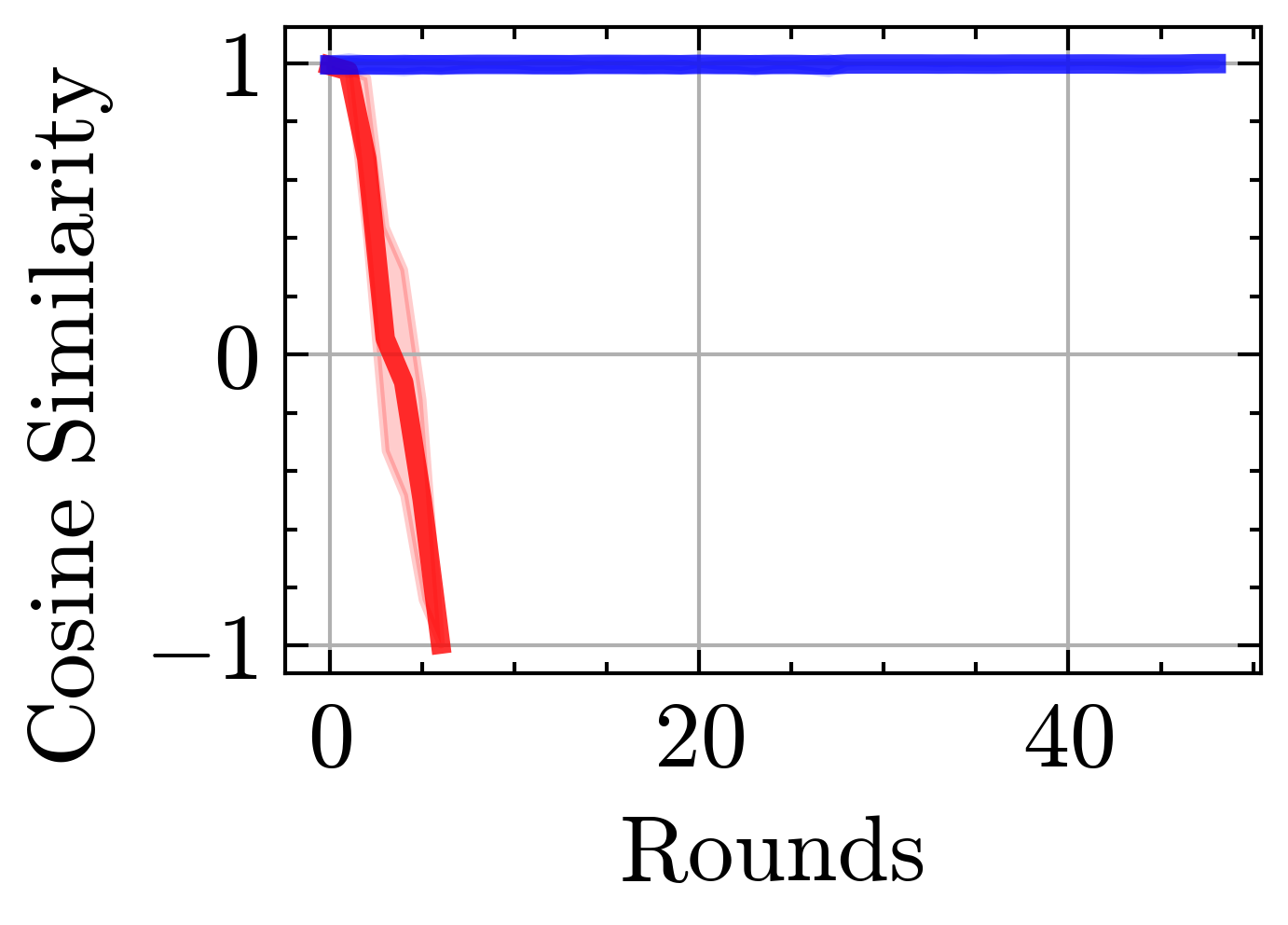}
	}
	\subfigure{
		\label{fig:image6}
		\includegraphics[width=.21\textwidth]{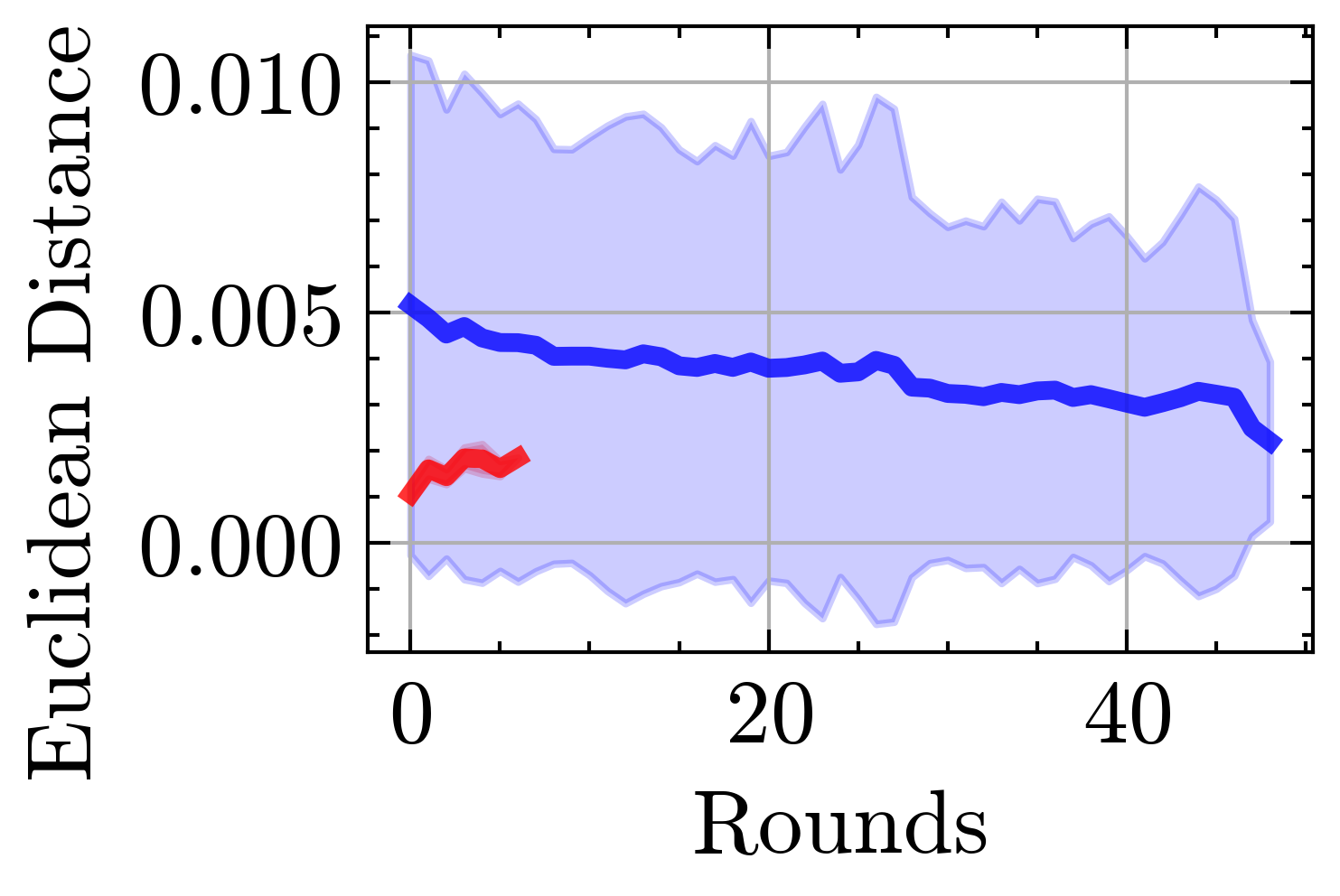}
	}
	\vspace{-1em}
	
	\addtocounter{subfigure}{-3}
	\subfigure[\textit{Min-Sum}~(Unit-Vector) attack]{
		\label{fig:PerfminSumAttack}
		\includegraphics[trim= 1500mm 0mm 0mm 0mm, clip, clip,width=.4\textwidth]{images/legendLocalBehave1.png}
	}\vspace{-0.5em}
	
	\subfigure{
		\label{fig:image7}
		\includegraphics[width=.21\textwidth]{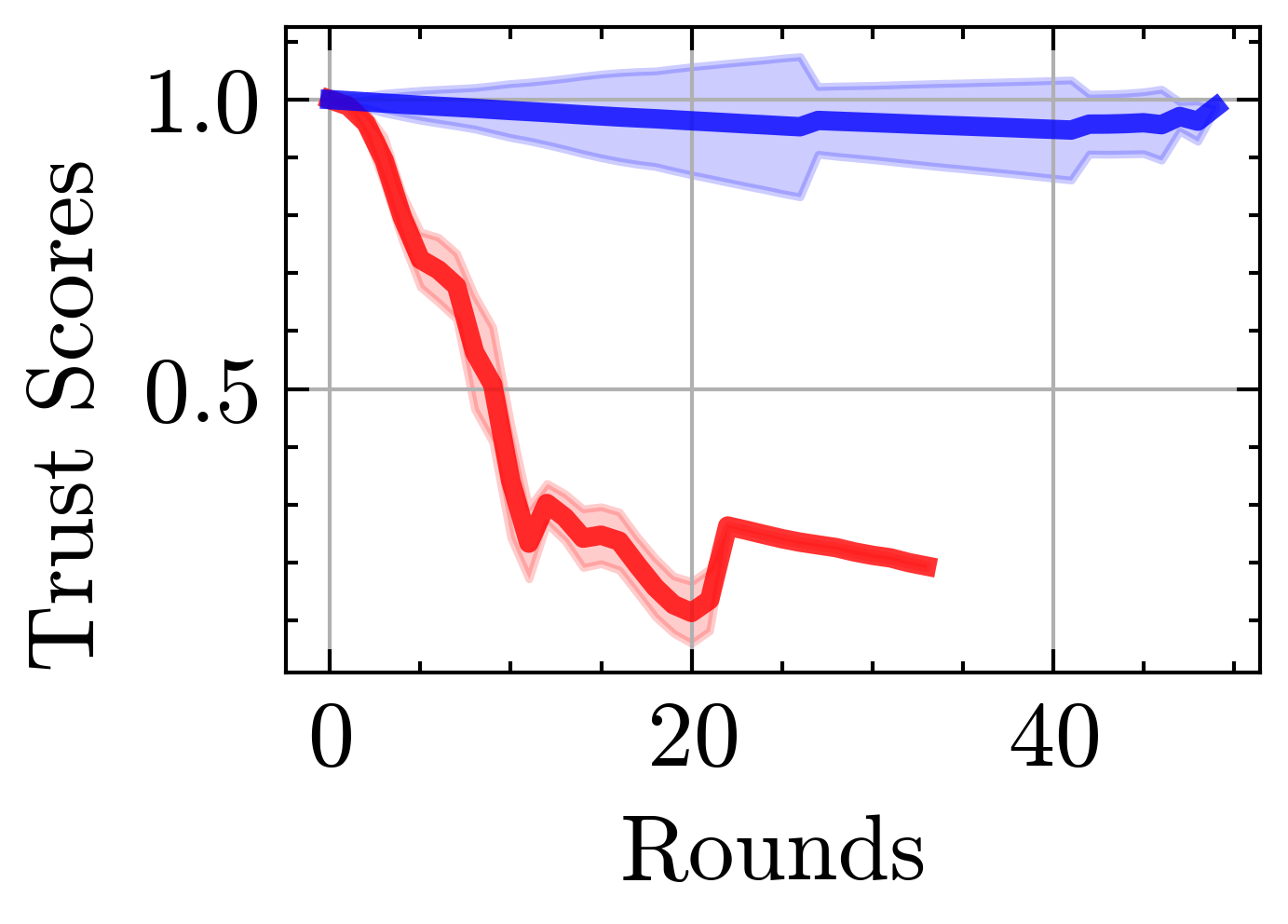}
	}
	\subfigure{
		\label{fig:image8}
		\centering
		\includegraphics[width=.21\textwidth]{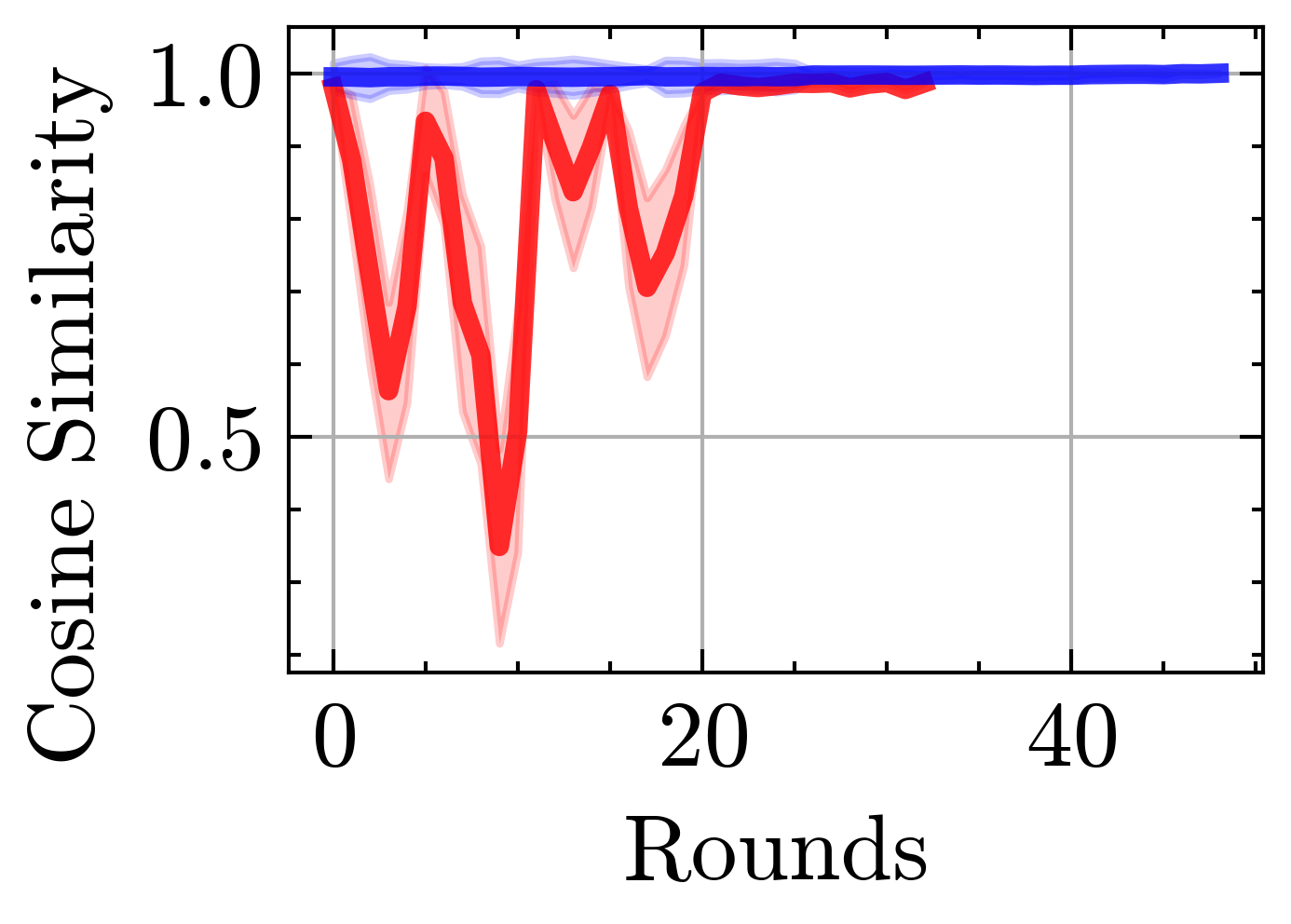}
	}
	\subfigure{
		\label{fig:image9}
		\includegraphics[width=.21\textwidth]{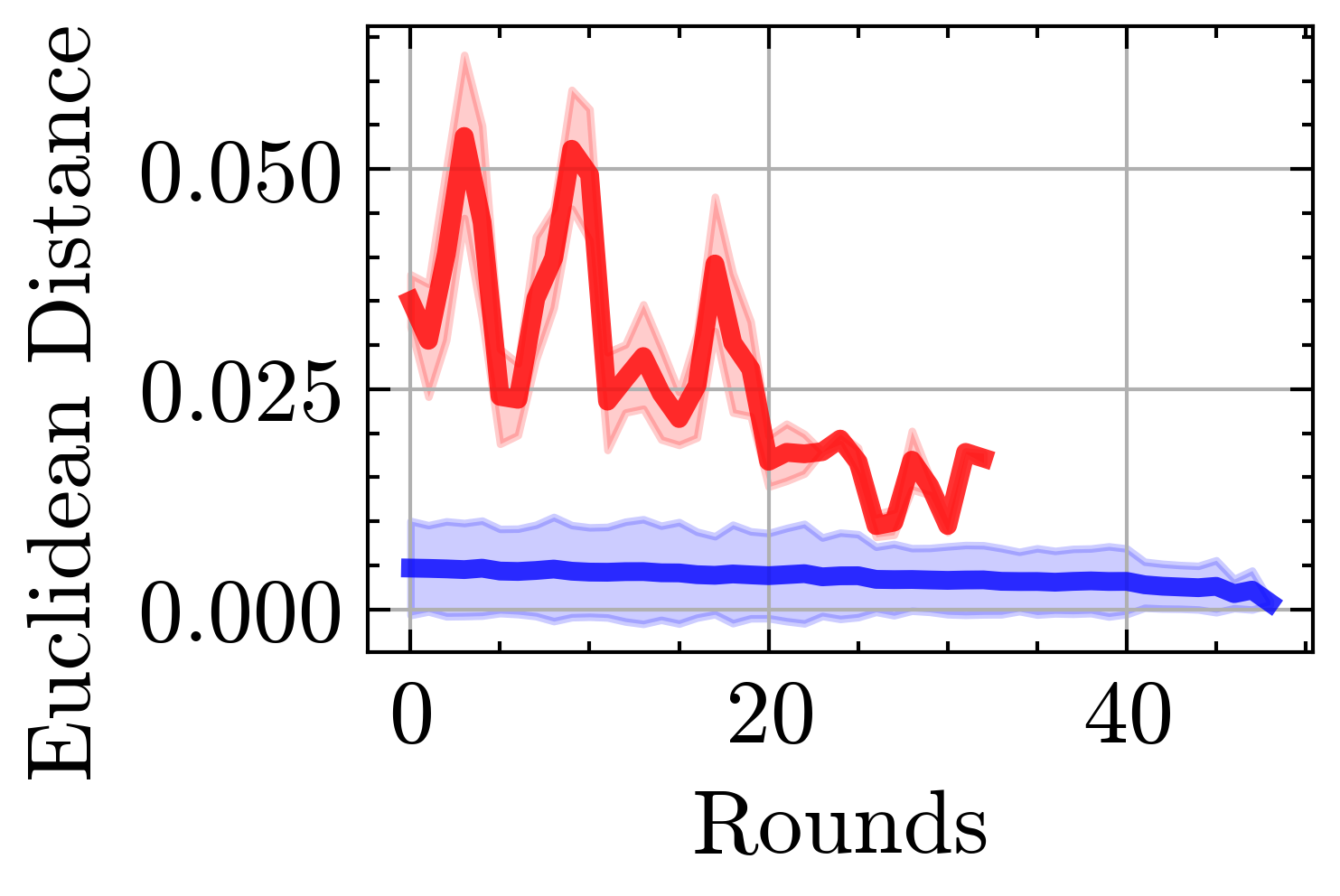}
	}
	\vspace{-1em}
	
	\addtocounter{subfigure}{-3}
	\subfigure[\textit{Krum-Attack}]{
		\label{fig:PerfKrumAttack}
		\includegraphics[trim= 1500mm 0mm 0mm 0mm, clip, clip,width=.3\textwidth]{images/legendLocalBehave1.png}
	}\vspace{-0.5em}
	
	\subfigure{
		\label{fig:image10}
		\includegraphics[width=.21\textwidth]{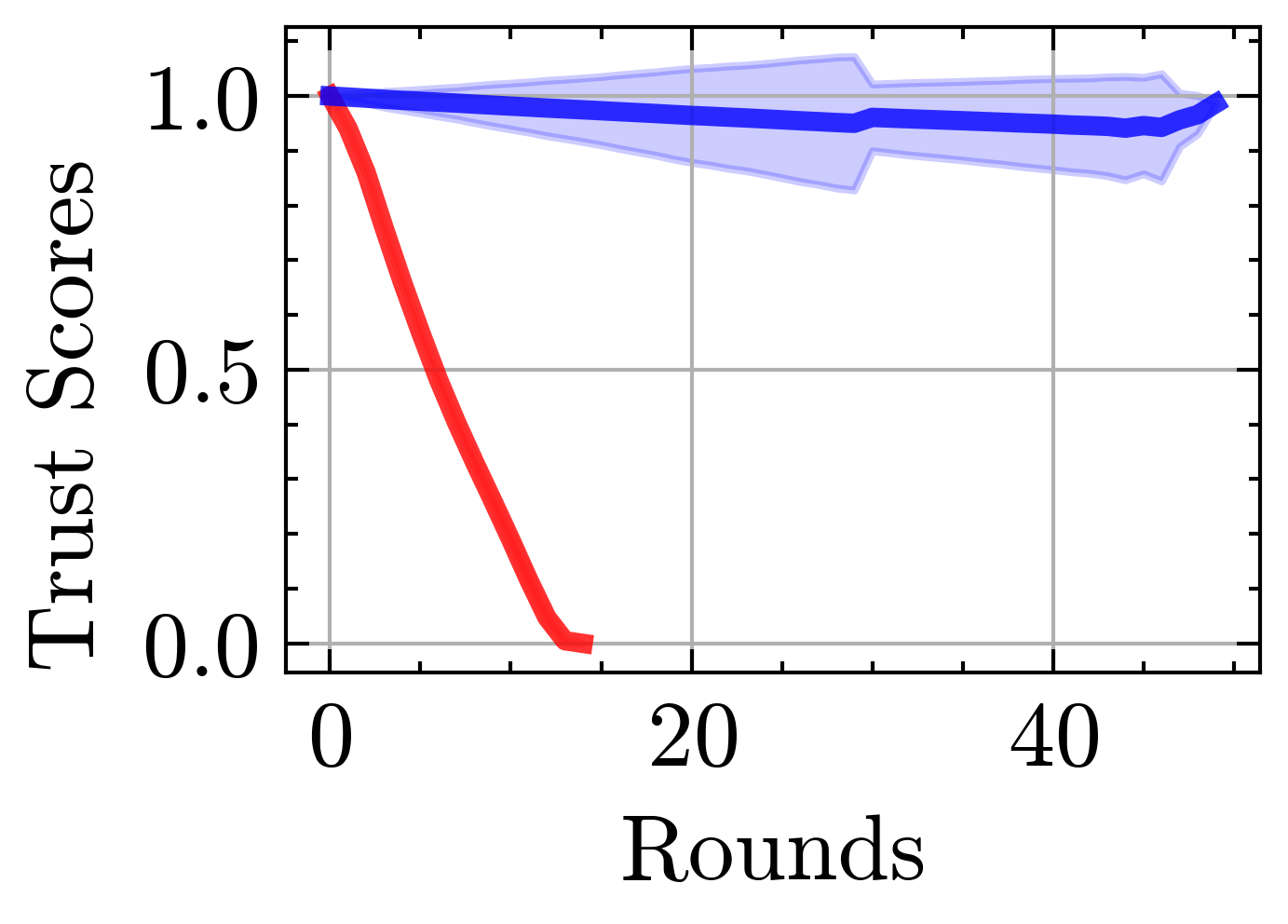}
	}
	\subfigure{
		\label{fig:image11}
		\includegraphics[width=.21\textwidth]{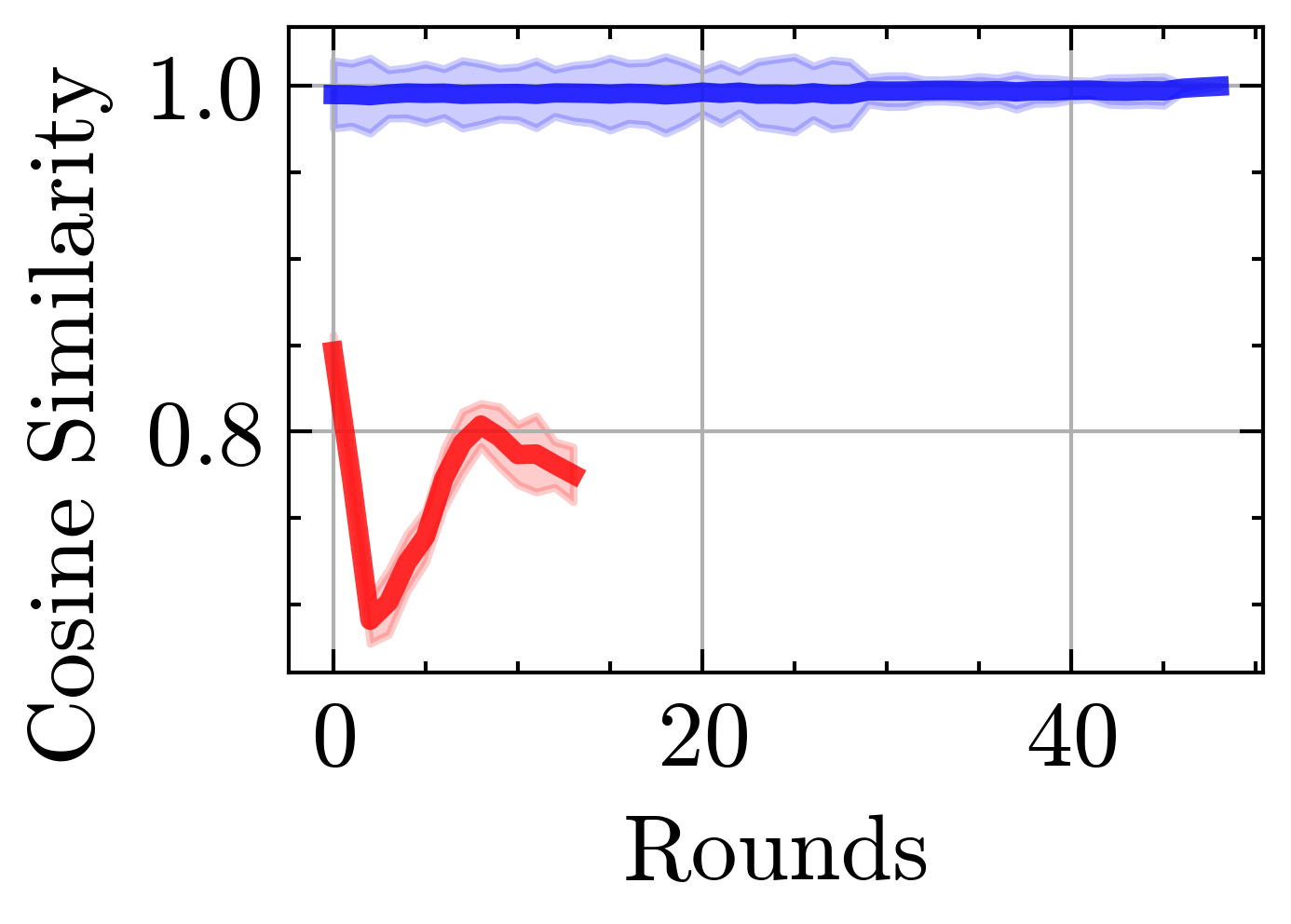}
	}
	\subfigure{
		\label{fig:image12}
		\includegraphics[width=.21\textwidth]{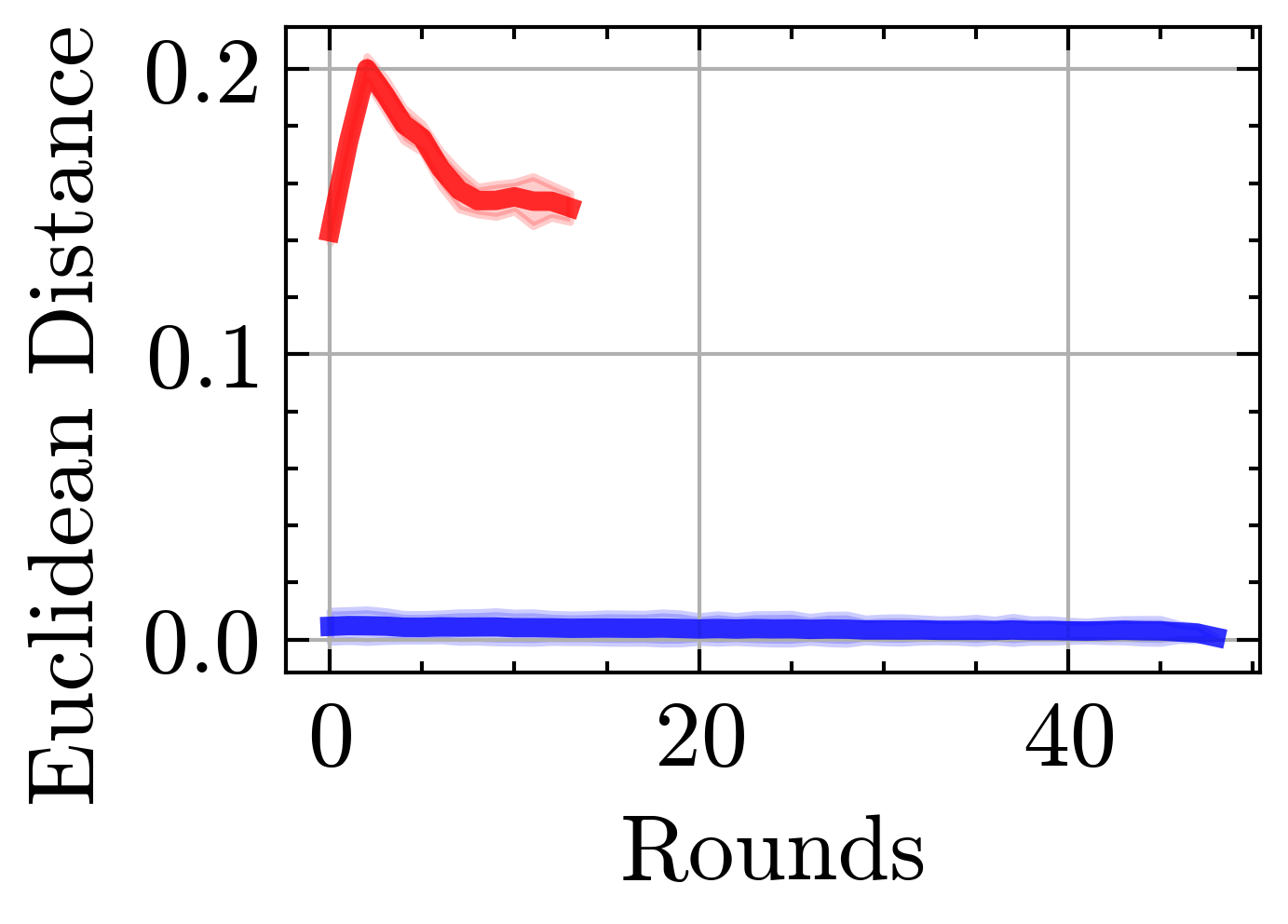}      
	}
	\vspace{-1em}
	
	\addtocounter{subfigure}{-3}
	\subfigure[\textit{Trim-Attack}]{
		\label{fig:PerfTrimAttack}
		\includegraphics[trim= 1500mm 0mm 0mm 0mm, clip, clip,width=.3\textwidth]{images/legendLocalBehave1.png}
	}
	\vspace{-1em}	
	\caption{Performance of \textit{KeTS} under different attacks.}
	\vspace{-1.3em}
	\label{fig:Perf}
\end{figure}

We now evaluate the behavior of malicious clients in the context of four optimization and aggregation-tailored attacks, i.e., \textit{Min-Max}~(cf. \figurename~\ref{fig:PerfminMaxAttack}), \textit{Min-Sum}~(cf. \figurename~\ref{fig:PerfminSumAttack}), \textit{Krum-Attack}~(cf. \figurename~\ref{fig:PerfKrumAttack}), and \textit{Trim-Attack}~(cf. \figurename~\ref{fig:PerfTrimAttack}). These scenarios are essential for: (1)~showcasing how attackers behave in a distinct and diverse manner compared to honest clients, and (2)~demonstrate the effectiveness of \textit{KeTS} for detection and mitigation.
\par
For each attack, we analyze the evolution of trust scores for both benign and malicious clients. Additionally, we examine the changes in cosine similarity and Euclidean distance. Cosine similarity~(cf.~Eq.~\ref{Eq:CS}) ranges from $-1$ to $1$, where $1$ signifies vectors pointing in the same direction and $-1$ indicates vectors pointing in opposite directions. Euclidean distance~(cf.~Eq.~\ref{Eq:ED}) accounts for the ${l}_{2}$ norm between consecutive updates, which helps in detecting modifications in the magnitude of the update that can amplify the poisoning effect. Each curve in \figurename~\ref{fig:Perf} represents the mean values for the metrics while shaded regions indicate standard deviations across clients sampled in a given round. Please note that malicious clients are excluded from sampling once their trust scores reach zero, which is marked by the disappearance of the red curves. Moreover, we note that the trust scores can increase between rounds, even though the formula should not allow this. This occurs because different clients may be sampled in different rounds, leading to variations in the mean value of the trust scores.
\par
As illustrated in \figurename~\ref{fig:Perf}, it is clear that malicious clients exhibit distinct behavior compared to benign clients. The trust scores for malicious clients in all four attacks sharply declines. Malicious clients experience drastic changes in update directions, as indicated by cosine similarity dropping below zero or its erratic fluctuation. The Euclidean distance for malicious clients also tends to drift away from zero~(in the case of the \textit{Min-Sum} attack, \textit{KeTS} quickly penalizes attackers based on the other two metrics). On the contrary, benign clients exhibit consistent behavior under all four attacks. In particular, benign clients maintain a stable trust score, cosine similarity near $1$, and the Euclidean distance near zero. To summarize, \textit{KeTS} is capable of identifying attackers in both optimization and aggregation-tailored attacks due to their distinct behavior.
\par
For our further investigations~(i.e., \S~\ref{noniidsValues} onwards), we present results for the MNIST dataset only; mainly due to the page limit, and the results for other datasets are congruent. We chose \textit{Min-Max}~(Unit-Vector) attack in these experiments because the trust score for malicious clients declines gradually~(cf.~\figurename~\ref{fig:Perf}) as compared to other attacks. Thus, it much more difficult to defend against. All other settings remain the same as described in~\S~\ref{configurations}, unless specified otherwise.

\subsection{Effect of degree of non-IIDnes}
\label{noniidsValues}
\begin{figure}[H]
	\vspace{-1.2em}
	\centering
	\includegraphics[width=0.3\textwidth]{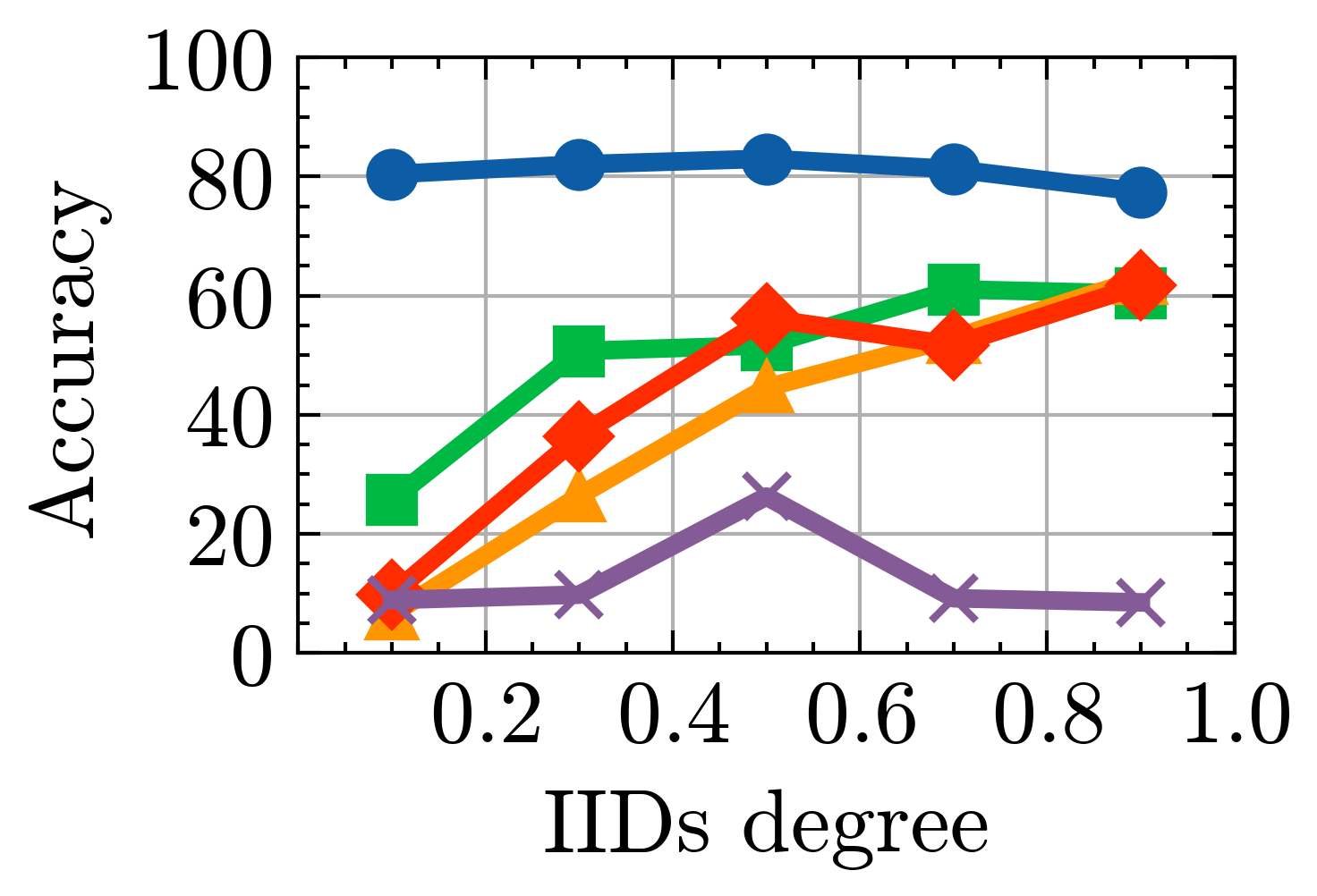}
	\includegraphics[trim= 0mm -18mm 0mm 0mm, width=0.4\textwidth]{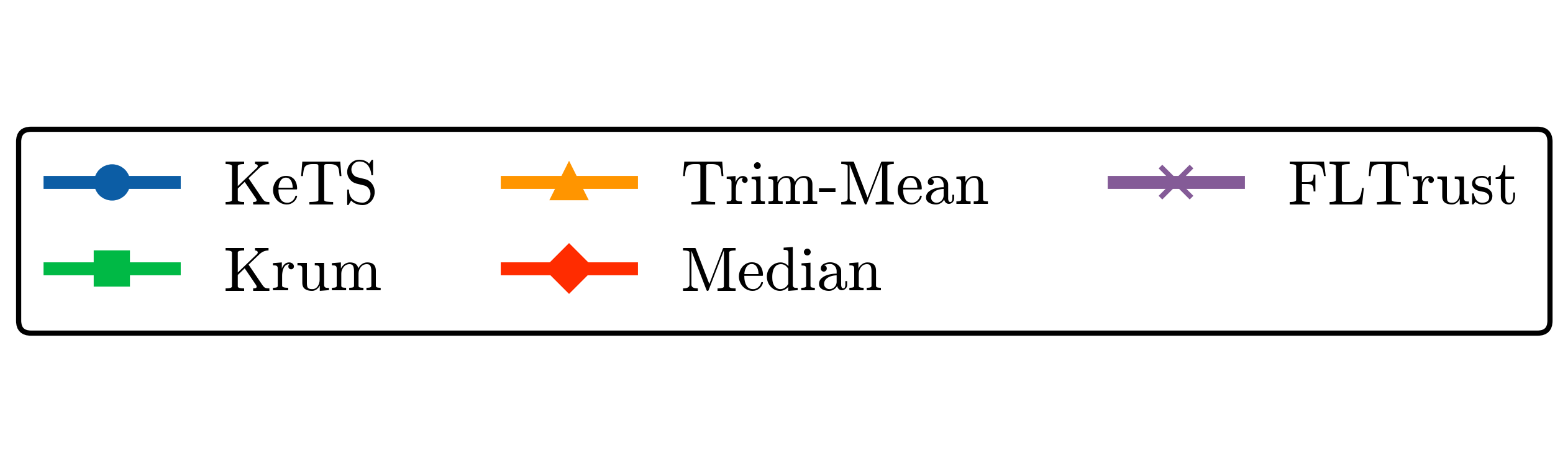}
	\vspace{-1em}
	\caption{Detection accuracy with respect to the degree of non-IIDness.}
	\vspace{-1em}
	\label{fig:nonIID}
\end{figure}

Next, we aim to analyze how \textit{KeTS} and classical defenses respond to modifications in the \textit{Dirichlet} factor, which influences the level of non-IIDness in the class distribution among the clients. Specifically, a lower value indicates a higher degree of non-IIDness in the class distributions. A higher degree of non-IIDness enables adversaries to craft more malicious gradients without detection, thereby amplifying the impact of their attacks. \figurename~\ref{fig:nonIID} presents the detection accuracy of different defense mechanisms concerning the degree of non-IIDness.
\par
It is evident from \figurename~\ref{fig:nonIID} that the majority of defense schemes perform better when the class distribution among clients becomes more balanced~(i.e., with a higher \textit{Dirichlet} factor). It is because the more the class samples are distributed in an IID manner, the fewer benign outliers there will be. \textit{FLTrust} demonstrates consistent yet suboptimal performance. This happens primarily due to the root dataset being too small to effectively represent the entire dataset; particularly in extreme non-IID environments, where it significantly deviates from the client's local data distributions. The accuracy in \textit{Krum}, \textit{Trim-Mean}, and \textit{Median} is directly proportional to the degree of IID-ness. Those defense techniques typically rely on statistical analyses of the update population, where the aggregation is computed based on the comparative information derived from multiple updates. However, such an approach is susceptible to failure in non-IID settings, where client data distributions exhibit significant disparities. In such scenarios, the presence of outliers can distort the aggregation results, even when robust measures such as the \textit{Median} or \textit{Trim-Mean} are applied.
\par
In contrast, \textit{KeTS} mitigates this issue by focusing on the individual update of each client. By evaluating updates independently, KeTS eliminates aggregation dependencies and reduces the influence of outliers. Our approach not only outperforms other defense techniques but also ensures stable accuracy across varying levels of non-IIDness.

\subsection{Impact of percentage of attackers}
\label{percentageOfAttackers}
\begin{figure}[H]
	\vspace{-1.2em}
	\centering
	\includegraphics[width=0.3\textwidth]{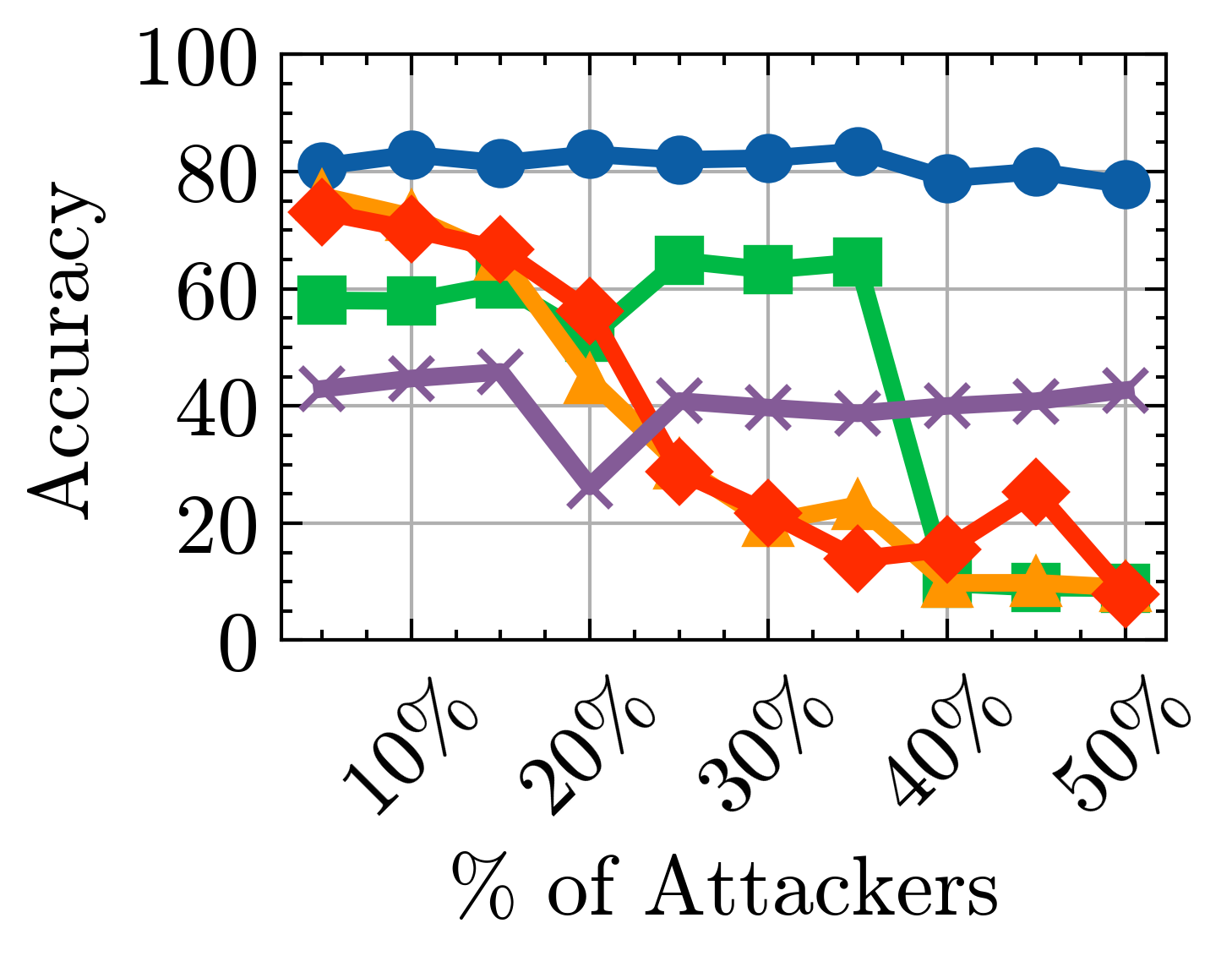}
	\includegraphics[trim= 0mm -25mm 0mm 0mm, width=0.4\textwidth]{images/legend_plotKeTS.png}
	\vspace{-1em}
	\caption{Accuracy on the server test set with respect to the percentage of attackers.}
	\vspace{-1em}
	\label{fig:perAttack}
\end{figure}

We also evaluate the impact of the attackers' population on detection accuracy. As shown in \figurename~\ref{fig:perAttack}, existing defense methods experience a substantial decline in performance as the number of attackers increases. For the same reason discussed in \S~\ref{noniidsValues}, this decline occurs because these methods rely on statistical analyses of the update population, where the aggregation is based on comparative information from multiple updates. As the number of attackers increases, the distribution of updates becomes increasingly skewed by malicious updates, impairing the effectiveness of these defenses.
In contrast, \textit{KeTS} focuses on analyzing each client’s update independently. Its performance remains consistent, with only minimal deterioration as the percentage of attackers increases.

\subsection{Different poisoning approaches}
\label{posioningApproaches}
\begin{figure}[H]
	\vspace{-1.2em}
	\centering
	\subfigure{
		\includegraphics[trim= 0mm 0mm 0mm 0mm, clip,width=.21\textwidth]{images/legendLocalBehave1.png}
	}
	\vspace{-1em}
	
	\subfigure{
		\includegraphics[width=.21\textwidth]{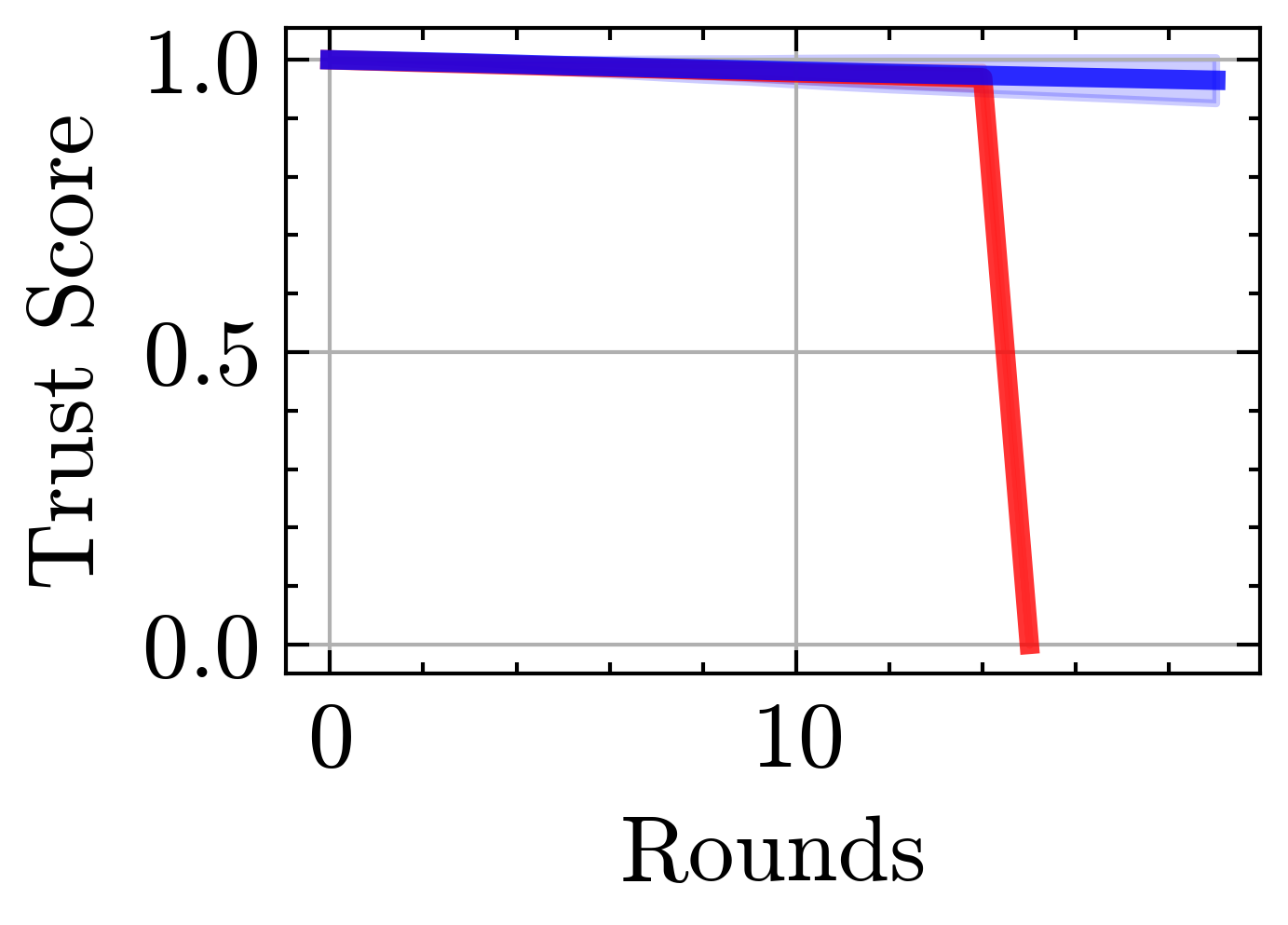}
	}
	\subfigure{
		\includegraphics[width=.21\textwidth]{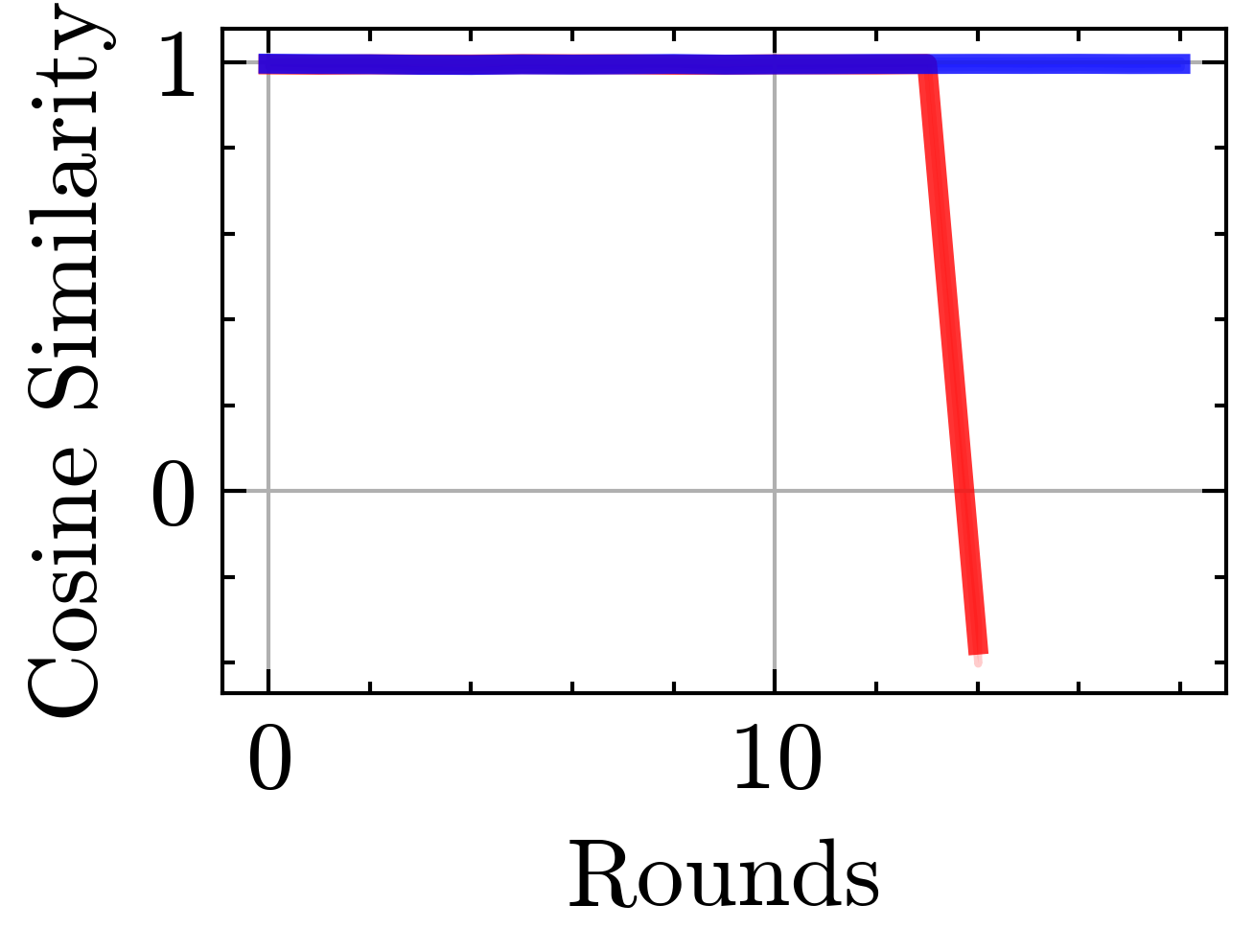}
	}
	\subfigure{
		\includegraphics[width=.23\textwidth]{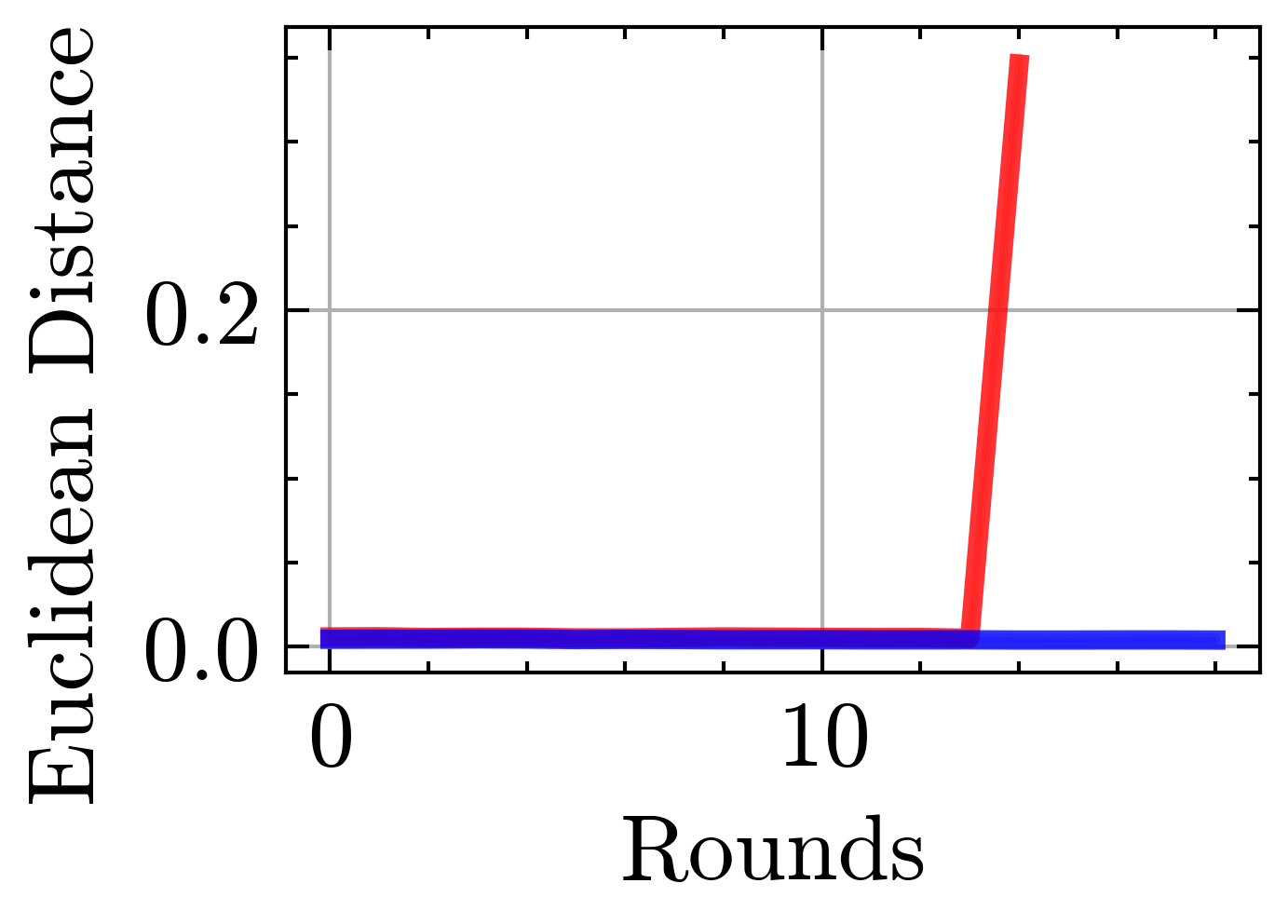}
	}
	\vspace{-1em}
	
	\addtocounter{subfigure}{-4}
	\subfigure[Start poisoning in between.]{
		\label{fig:startBetween}
		\includegraphics[trim= 1500mm 0mm 0mm 0mm, clip, clip,width=.4\textwidth]{images/legendLocalBehave1.png}
	}\vspace{-0.5em}
	
	\subfigure{
		\includegraphics[width=.21\textwidth]{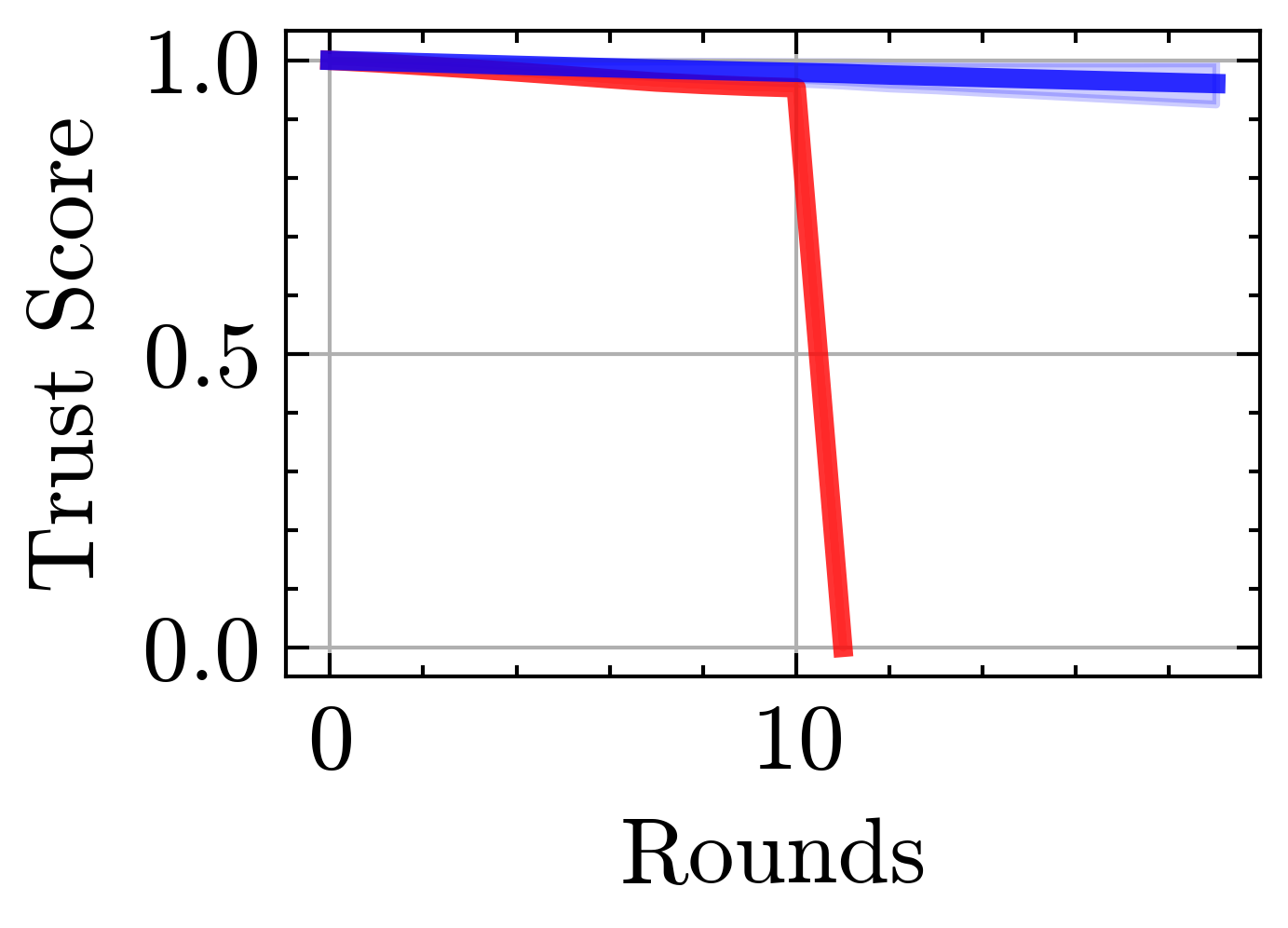}
	}
	\subfigure{
		\includegraphics[width=.21\textwidth]{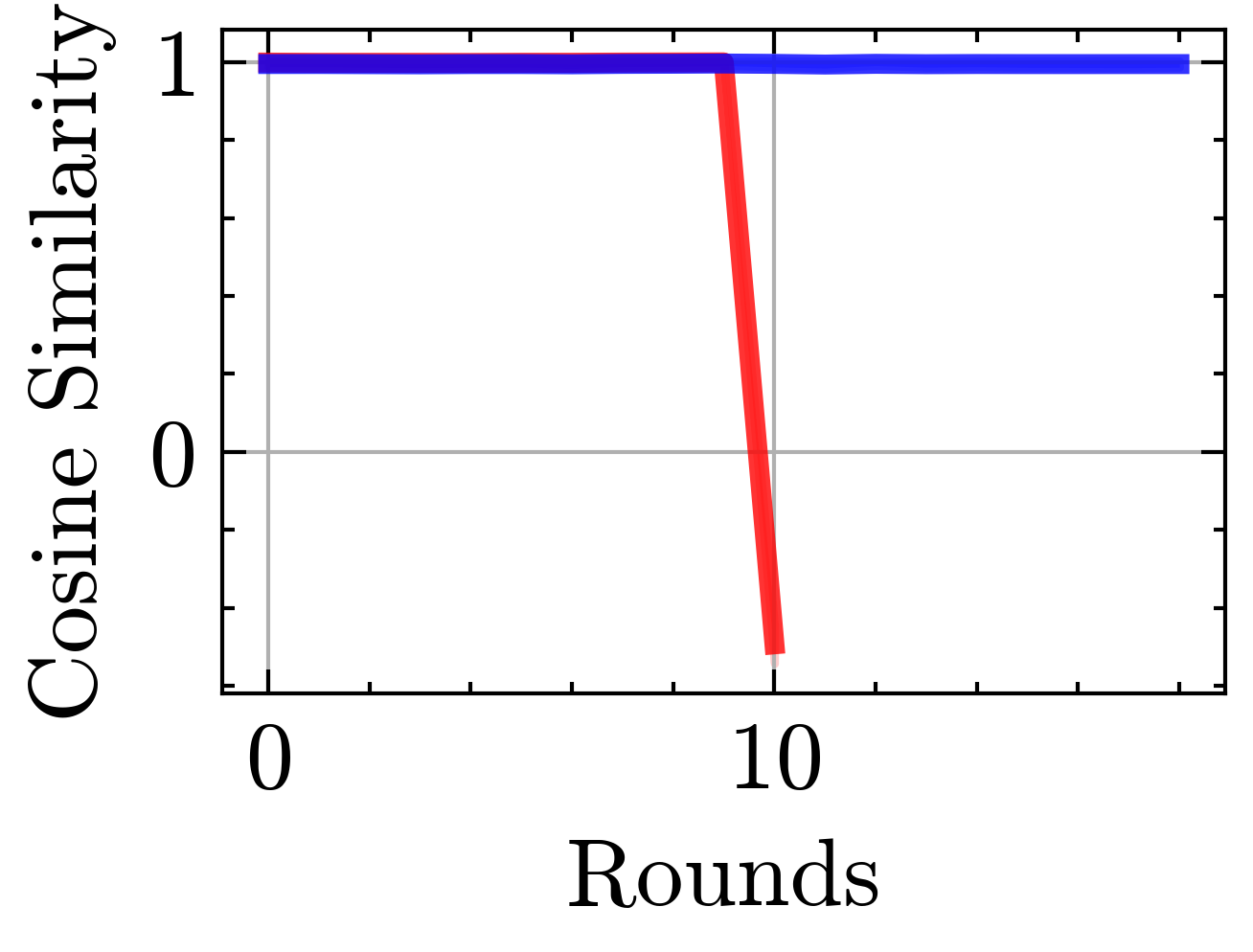}
	}
	\subfigure{
		\includegraphics[width=.21\textwidth]{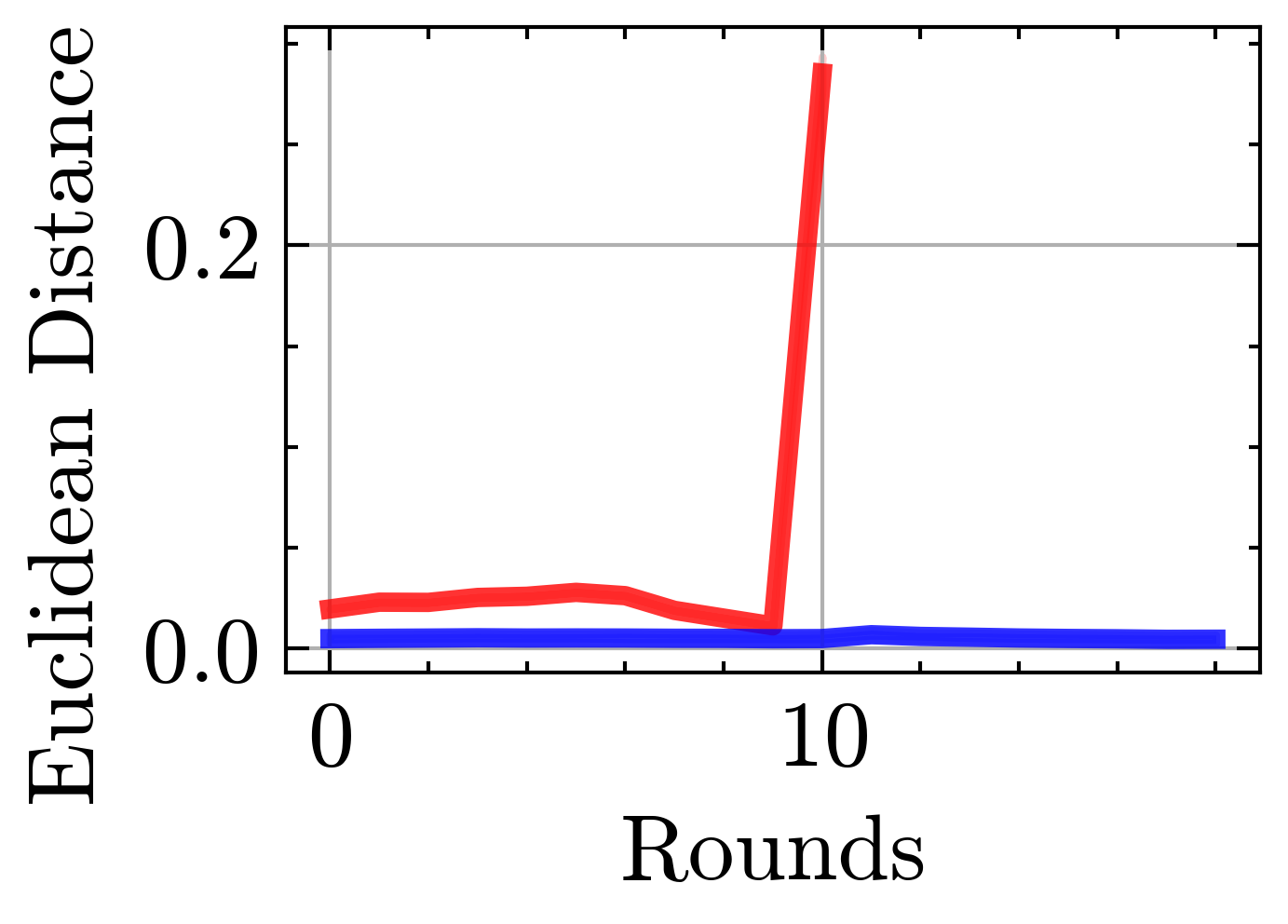}
	}
	\vspace{-1em}
	
	\addtocounter{subfigure}{-3}
	\subfigure[Stop poisoning in between.]{
		\label{fig:stopBetween}
		\includegraphics[trim= 1500mm 0mm 0mm 0mm, clip, clip,width=.4\textwidth]{images/legendLocalBehave1.png}
	}
	\vspace{-1em}
	\caption{Performance of \textit{KeTS} under different poisoning approaches.}
	\vspace{-1em}
\end{figure}

In all the previous experiments, the attacker consistently sent malicious updates starting from the first epoch. We now consider two different poisoning strategies. In the first approach, the attacker begins to send malicious updates once the FL training has already started, i.e., during a global epoch $>0$. \textit{KeTS} promptly identifies such attacks, as the injected uploads are highly inconsistent with the previous legitimate updates. Specifically, these uploads exhibit a cosine similarity~$< 0$, leading to an immediate assignment of a trust score of zero~(cf.~\figurename~\ref{fig:startBetween}).
\par
In the second approach, we analyze the scenario where an attacker ceases to send malicious updates and instead submits legitimate ones after training on its dataset. In this case, \textit{KeTS} detects the abrupt change in direction, identifies all the attackers, and sets their trust scores to~0~(cf.~\figurename~\ref{fig:stopBetween}).

\subsection{Impact of number of local epochs}
\label{numberOfLocalEpochs}
\begin{figure}[H]
	\vspace{-1.2em}
	\centering
	\includegraphics[trim= 0mm 0mm 0mm 5mm, clip, width=0.32\textwidth]{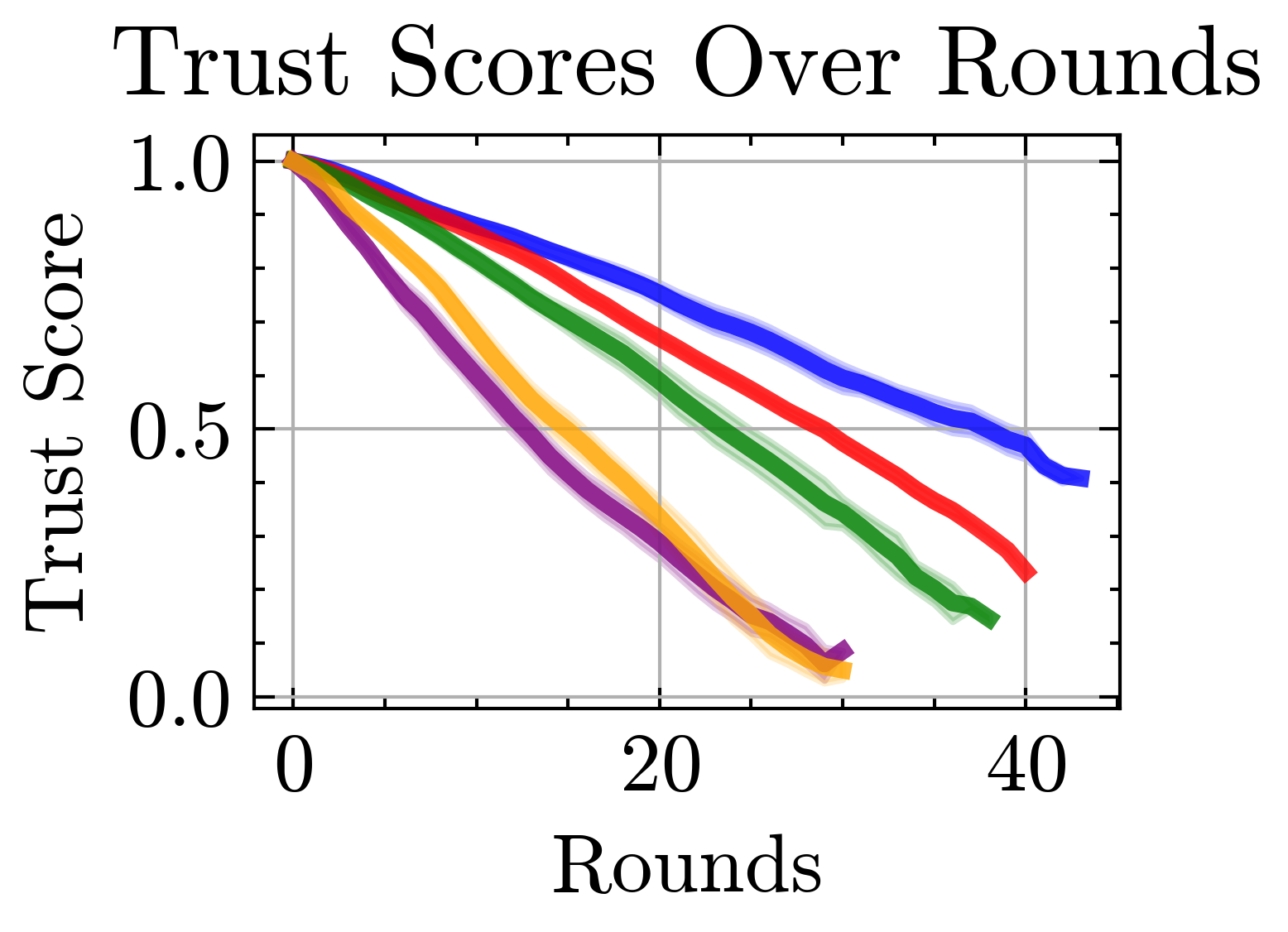}
	\includegraphics[trim= 0mm -18mm 0mm 0mm, width=0.2\textwidth]{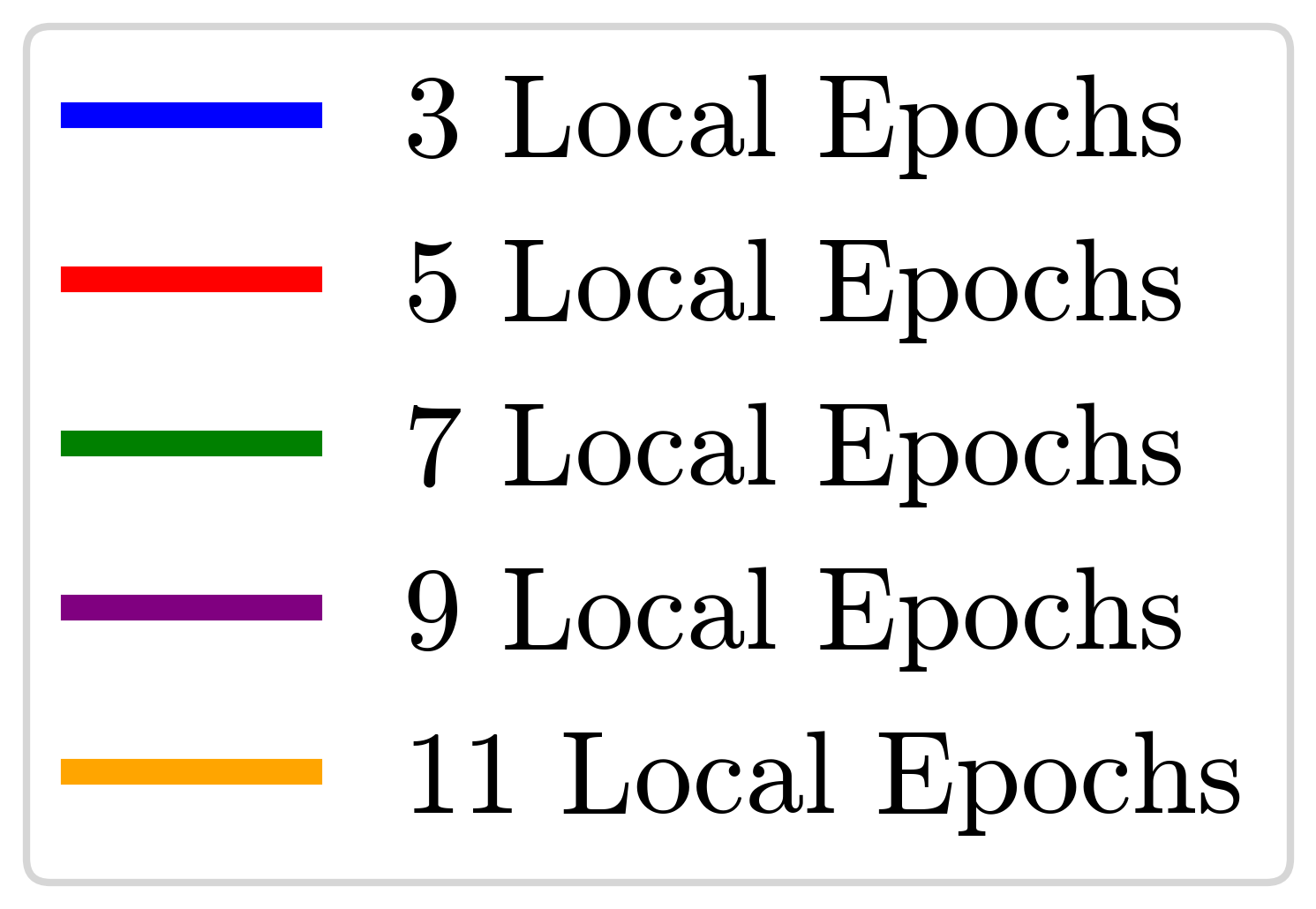}
	\vspace{-1em}
	\caption{Trust score decline for attackers in \textit{KeTS} over different numbers of local epochs.}
	\label{localEpochs}
	\vspace{-1em}
\end{figure}

We now investigate how changing the number of local epochs for each client influences the speed at which trust scores decrease for attackers. \figurename~\ref{localEpochs} shows that increasing the number of local epochs causes the attackers' trust scores to decrease more rapidly, resulting in their exclusion from sampling in a shorter amount of time. We hypothesize that this happens because, with more local epochs, the updates become larger and more  heterogeneous. It allows the attacker to have a bigger radius of search to find a malicious upload. As a result, the difference in direction and magnitude for the attacker becomes more pronounced.

\subsection{\textit{KeTS'} extension for consistent-untargeted and targeted attacks}
\label{sec:KeTSv2}
We now demonstrate that KeTS can be easily extended to counter other attacks that exhibit a more natural sequence of updates (i.e., consistent-untargeted and targeted attacks).  In particular, we examine the \textit{sign-flipping} attack~\cite{sharma2024probabilistic, Chelli2023FedGuardSP,li2019rsa}, where the attacker computes the update in a legitimate manner, after which the sign is flipped, resulting in consistent updates that point in an adversarial direction. Next, we consider the \textit{label-flipping} attack~\cite{biggio2012poisoning, fung2020limitations}, where labels are altered before the training process begins, causing the updates to follow a natural progression. We explore the following two solution avenues:
\par
1. Use KeTS as a preliminary step/filter before any aggregation scheme. Here, KeTS detects untargeted, inconsistent attacks (as discussed throughout this paper), after which any standard aggregation method can be applied to the client updates. This approach enhances the effectiveness of the existing defenses.
\par
2. A simple extension of KeTS, referred to as KeTSv2, handles attacks that misdirect updates to hinder convergence. Unlike crafted attacks that mimic benign behavior, these attacks rely on updates derived from legitimate training but oriented in a divergent direction. After filtering clients based on trust scores, KeTSv2 computes the cosine similarity between each client update and the previous global update, discarding those falling below a defined threshold. This method assumes that the number of attackers is less than half of the clients. The previous global update - used in the similarity check - is computed at each epoch using a momentum-based update rule:
$\theta_t = (1 - \mu) \cdot \theta_{t-1} + \mu \cdot update_{new}$.
\par
\tablename~\ref{tab:KeTSv2} shows our results on the final accuracy, based on the server test set, on Fashion-MNIST using $\mu = 0.1$, a Dirichlet distribution with concentration parameter $2.0$, over $80$ global rounds, $5$ local training rounds per client, and $80$ clients with full client participation at each round. These results showcase that KeTS is a complete solution against both targeted and untargeted attacks.
\begin{table}[H]
	\vspace{-1em}
	\centering
	\caption{KeTS (v1, as a pre-filter, v2) against \textit{sign-} and \textit{label-flipping} attacks. In pre-filter setting, we test KeTS + Median as Median is the best among existing defenses.}
	\label{tab:KeTSv2}
	\resizebox{\columnwidth}{!}{%
	\begin{tabular}{|c|cccc|c|c|c|}
		\hline
		\multirow{2}{*}{\textbf{Attacks}} & \multicolumn{4}{c|}{\textbf{Existing defenses}} & \multirow{2}{*}{\textbf{\begin{tabular}[c]{@{}c@{}}KeTS\\ (v1)\end{tabular}}} & \textbf{\begin{tabular}[c]{@{}c@{}}KeTS as a \\ pre-filter to \\ another defense\end{tabular}} & \multirow{2}{*}{\textbf{\begin{tabular}[c]{@{}c@{}}KeTS \\ (v2)\end{tabular}}} \\ \cline{2-5} \cline{7-7}
		& \multicolumn{1}{c|}{\textbf{Median}} & \multicolumn{1}{c|}{\textbf{FLTrust}} & \multicolumn{1}{c|}{\textbf{Trim-Mean}} & \textbf{Krum} &  & \textbf{KeTS+Median} &  \\ \hline
		\textit{Label-flipping} & \multicolumn{1}{c|}{69.18\%} & \multicolumn{1}{c|}{65.13\%} & \multicolumn{1}{c|}{69.09\%} & 66.90\% & 60.47\% & 69.28\% & 72.20\% \\ \hline
		\textit{Sign-flipping} & \multicolumn{1}{c|}{59.54\%} & \multicolumn{1}{c|}{65.07\%} & \multicolumn{1}{c|}{59.17\%} & 69.30\% & 30.00\% & 59.97\% & 73.16\% \\ \hline
	\end{tabular}
}
\vspace{-1em}
\end{table}

\section{Conclusion}
\label{conclusion}
In this paper, we propose \textit{KeTS} - a novel defense for FL against untargeted model poisoning attacks. We choose a direction based on the evaluation of each client individually. Our evaluations demonstrate that \textit{KeTS} outperforms existing defenses and effectively handles the honest outlier problem. Moreover, \textit{KeTS} is a scalable solution because it incurs minimal computation and storage overheads on the server and imposes no additional overhead on the clients.

\bibliographystyle{splncs04}
\bibliography{bibS}
\balance

\begin{subappendices}
	\renewcommand{\thesection}{\Alph{section}}%
	\counterwithin{table}{section}
	\counterwithin{figure}{section}
	\counterwithin{algorithm}{section}
	\section{Algorithms}
	\label{appendix:algo}
	\vspace{-3em}
	\addtocounter{algorithm}{-1}
		\begin{algorithm}[H]
			\label{pseudo1}
			\SetKwData{Left}{left}
			\SetKwData{This}{this}
			\SetKwData{Up}{up}
			\SetKwFunction{Union}{Union}
			\SetKwFunction{FindCompress}{FindCompress}
			\SetKwInOut{Input}{input}
			\SetKwInOut{Output}{output}
			\caption{KDE}
			\Input{$TScores$ (list of trust scores),
				$SampledClients$ (list of sampled clients id),
				$EstimateBandwidth$ (function to estimate the bandwidth).
			}
			\Output{$HonestClients$ list of honest Clients}
			\BlankLine
			$bandwidth \gets \texttt{EstimateBandwidth}(TScores)$\;\tcp*{get the kernel bandwidth estimation}

			$density \gets \texttt{EstimateDensity}(TScores, bandwidth)$\;\tcp*{Generate the density estimation}
			
			$x_d \gets Generate\ density\ values\ in\ range\ (0,\texttt{max}(TScores)+1)$\;

			$MinimaIndices \gets \texttt{FindLocalMinima}(density)$\;\tcp*{Indexes of local minima of density function}
			
			$ClusterBoundaries \gets x_d[MinimaIndices]$\;\tcp*{Trust scores that mark the beginning of segments}

			$HonestSegment \gets ClusterBoundaries[-1]$\;\tcp*{Index of the Trust Score marking the start of the segment}
			
			\For{client $c \in  SampledClients$}{
				\If{$TScores[c] >= HonestSegment$}{
					Append $c$ to $HonestClients$
				}
			}
			\caption{KeTS-KDE-Segmentation}
		\end{algorithm}
	\setcounter{AlgoLine}{0}
	\begin{algorithm}[H]
		\label{pseudo2}
		\SetKwData{Left}{left}
		\SetKwData{This}{this}
		\SetKwData{Up}{up}
		\SetKwFunction{Union}{Union}
		\SetKwFunction{FindCompress}{FindCompress}
		\SetKwInOut{Input}{input}
		\SetKwInOut{Output}{output}
		\caption{KeTS-Aggregation}
		\Input{$TScores$ (Trust scores),
			$\beta$ (constant),
                $Py$ (penalty),
			$G$ (set of uploaded updates by sampled clients), 
			$SampledClients$ (list of sampled clients id),
			$ClientsUpdates$ (list of previous updates for each client),
			$w$ (no. of samples in each client dataset).
		}
		\Output{New aggregated update}
		\BlankLine
		\ForEach{client $c \in SampledClients$}{
			$sim \gets \texttt{CosineSimilarity}(client\_updates[c][-1], G[c])$\;
			
			$eudist \gets \texttt{Euclidean distance}(ClientsUpdates[c][-1], G[c])$\;
			
			\eIf{$sim < 0$}{
				$TScores[c] \gets 0$\;
			}{
				$Py \gets (1 - sim) + eudist$\;
				
				$Tscores[c] \gets \max(0, TScores[c] - \beta \cdot Py)$\;
			}
			
		}
		\BlankLine
		$HonestClients \gets \texttt{KDESegmentation}(TScores, SampledClients)$\;
		
		$TotalSamples \gets \sum_{i \in \text{HonestClients}} w[i]$
		
		$AggregatedUpdate \gets \sum_{i \in\text{HonestClients}}\frac{w[i]}{TotalSamples}ClientUpdates[i]$\;\tcp*{Same aggregation as \textit{FedAvg} but with updates rather then models. The resulting update will be added to the previous global model.}
	\end{algorithm}

\section{Can \textit{FLTrust} match \textit{KeTS}'s performance?}
\label{fltrust}
\begin{figure}[H]
	\vspace{-1.6em}
	\centering
	\includegraphics[width=0.31\textwidth]{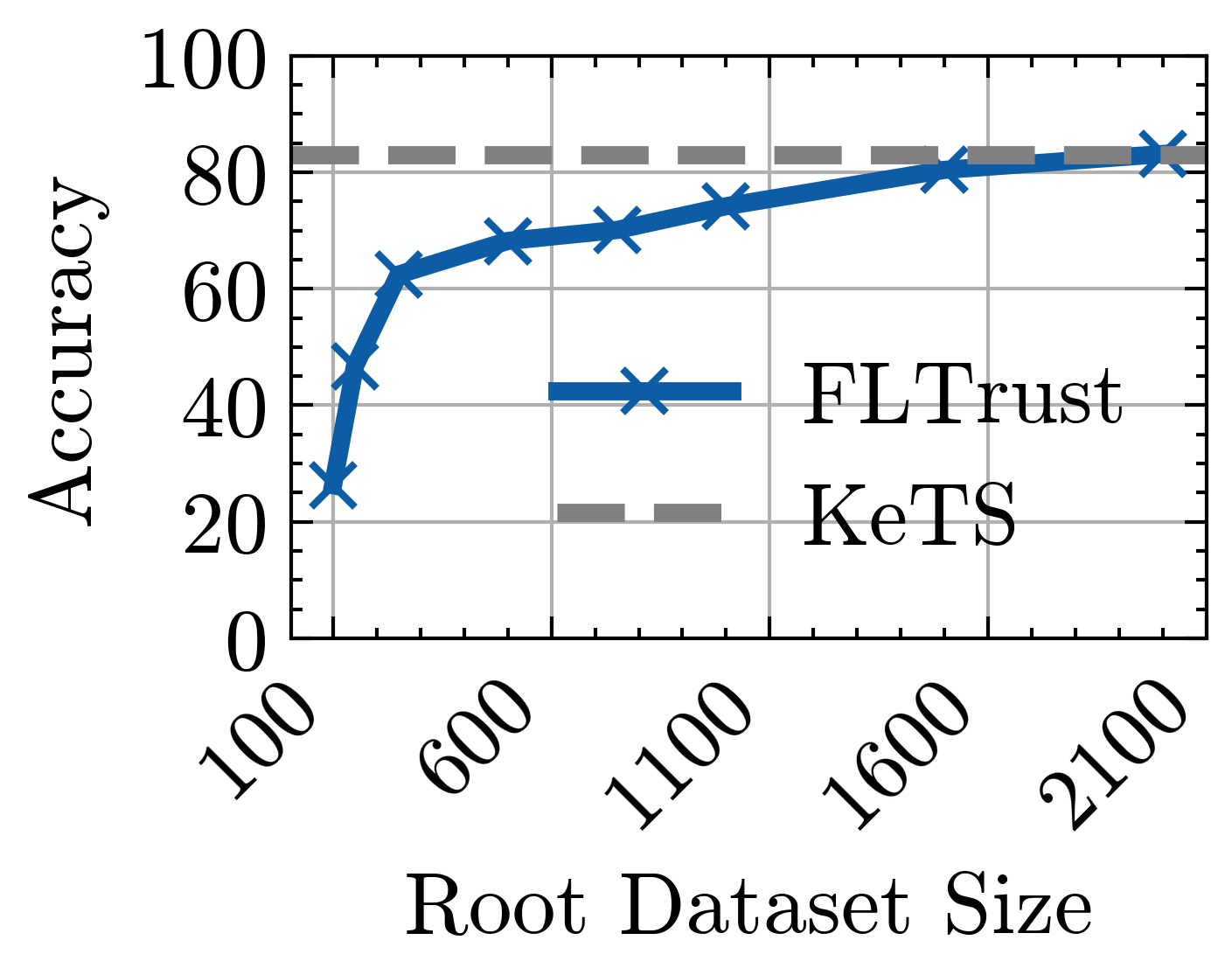}
	\vspace{-1em}
	\caption{Comparison with \textit{FLTrust}.}
	\label{flrootsize}
\end{figure}

We now analyze if \textit{FLTrust}, a state-of-the-art defense, can achieve the same performance as \textit{KeTS} under the same settings as those used in \figurename~\ref{fig:MNIST}. By increasing the size of the root dataset, the performance of \textit{FLTrust} improves~(cf.~\figurename~\ref{flrootsize}). \textit{FLTrust} requires over 2000 samples~(which is over three times the size of a client's dataset in this setting; cf.~\tablename~\ref{tab:setting}) in the root dataset to match the performance with \textit{KeTS}. This requirement poses significant challenges for the server and is practically infeasible\footnote{Due to strict privacy constraints, the server is prohibited from directly accessing client data, preventing it from sampling client data to estimate their distribution. As a result, the server must rely on externally sourced datasets. This limitation is further exacerbated in non-IID settings, where clients often hold only a subset of classes.} in FL.
\end{subappendices}	
\end{document}